\documentclass[prl,twocolumn,preprintnumbers,amsmath,amssymb,superscriptaddress]{revtex4-1}

\usepackage[ansinew]{inputenc}
\usepackage[T1]{fontenc}
\usepackage{ae,aecompl}
\usepackage[english]{babel}
\usepackage[dvips]{graphicx}
\usepackage{dsfont}
\usepackage{amsmath}
\usepackage{pifont}
\usepackage{enumerate,color,float}

\usepackage[caption=false]{subfig}

\newcommand{\braket}[2]{\mbox{$\langle #1|#2\rangle$}}

\newcommand{\op}[1]{\mbox{$\hat{#1}$}}
\newcommand{\ket}[1]{\vert#1\rangle}
\newcommand{\bra}[1]{\langle#1\vert}
\newcommand{\ud}{\,\mathrm{d}}
\newcommand{\amp}[1]{\mbox{\boldmath $\mathrm{#1}$}}

\begin{document}

\title{Channel purification via continuous-variable quantum teleportation with Gaussian post-selection}
\author{R\'emi Blandino}
\email{r.blandino@uq.edu.au}
\affiliation{Centre for Quantum Computation and Communication Technology, School of Mathematics and Physics, University of Queensland, St Lucia, Queensland 4072, Australia}
\author{Nathan Walk}
\affiliation{Centre for Quantum Computation and Communication Technology, School of Mathematics and Physics, University of Queensland, St Lucia, Queensland 4072, Australia}
\author{Austin P. Lund}
\affiliation{Centre for Quantum Computation and Communication Technology, School of Mathematics and Physics, University of Queensland, St Lucia, Queensland 4072, Australia}
 \author{Timothy C. Ralph} 
 \email{ralph@physics.uq.edu.au}
 \affiliation{Centre for Quantum Computation and Communication Technology, School of Mathematics and Physics, University of Queensland, St Lucia, Queensland 4072, Australia}

\begin{abstract}
We present a protocol based on continuous-variable quantum teleportation and Gaussian post-selection that can be used to correct errors introduced by a lossy channel. We first show that the global transformation enacted by the protocol is equivalent to an effective system composed of a noiseless amplification (or attenuation), and an effective quantum channel, which can in theory have no loss and an amount of thermal noise arbitrarily small, hence tending to an identity channel. An application of our protocol is the probabilistic purification of quantum non-Gaussian states using only Gaussian operations. 

\end{abstract}
\maketitle


\paragraph{Introduction} Practical quantum communication protocols such as quantum teleportation \cite{bennett_teleporting_1993,braunstein_teleportation_1998,bouwmeester_experimental_1997,boschi_experimental_1998,
furusawa_unconditional_1998,bowen_experimental_2004,takei_experimental_2005,lee_teleportation_2011} or Quantum Key Distribution (QKD) \cite{scarani_security_2009}, rely on the ability to transmit a quantum state over an imperfect quantum channel. Loss, as a main source of error, introduces Gaussian noise which cannot be compensated by deterministic amplification \cite{caves_quantum_1982}. Certain no-go theorems \cite{niset_no-go_2009,eisert_distilling_2002,fiurasek_gaussian_2002,giedke_characterization_2002} also restrict the ability of non-deterministic Gaussian operations to help and has led to the general belief that their utility is strictly limited. This is unfortunate because such operations are ubiquitous, since Gaussian states are relatively easy to produce and require simple detectors to measure their properties \cite{weedbrook_gaussian_2012}. 

While standard deterministic Gaussian operations are of no help, loss can however be corrected using a Noiseless Linear Amplifier (NLA) in a quantum teleportation scheme \cite{ralph_quantum_2011}, or in addition to using a noiseless linear attenuator applied before the quantum channel \cite{micuda_noiseless_2012}.  The noiseless linear amplifier (resp. attenuator) is a probabilistic operation described by an unbounded operator $g^{\hat{n}}$ \footnote{Note that $g^{\hat{n}}$ is actually unbounded only for $g{>}1$.}, with $g{\geq}1$ (resp. $g{\leq}1$), which transforms a coherent state $\ket{\alpha}$ to \cite{ralph_nondeterministic_2008,fiurasek_engineering_2009,marek_coherent-state_2010}
\begin{align}
g^{\hat{n}}\ket{\alpha} = e^{\frac{1}{2}(g^2{-}1)\vert \alpha \vert^2}\ket{g \alpha}.
\label{tranformation_NLA}
\end{align}
The NLA features many interesting properties and transforms mixed states in a non-trivial way \cite{walk_nondeterministic_2013,blandino_noiseless_2014}, which can result in a reduction of the encountered loss. It has been shown to be useful for several applications, such as continuous-variable QKD \cite{blandino_improving_2012,walk_security_2013,fiurasek_gaussian_2012}.  While it is in theory a Gaussian operation, its unboundedness prevents a perfect implementation, which, as a result, will
be only approximately Gaussian. Most of the experimental implementations are indeed based on non-Gaussian resources such as single-photon ancilla and photon counting \cite{ferreyrol_implementation_2010,zavatta_high-fidelity_2011,xiang_heralded_2010}, which make a practical use technically challenging.

When a measurement is involved in a quantum information protocol, post-selection is a convenient way to improve performances by keeping only certain outcomes. A simple example is the retention of only certain `click' patterns in single photon experiments. Post-selection can also be employed in continuous-variable experiments, for instance in continuous-variable QKD \cite{silberhorn_continuous_2002}, or for the teleportation of a non-Gaussian Wigner function \cite{mista_continuous-variable_2010}. Interestingly, it has been shown recently that the transformation obtained using an NLA immediately before an heterodyne measurement can be reproduced by post-selection, without the need of an actual physical implementation of the NLA \cite{fiurasek_gaussian_2012}. One simply needs to keep an heterodyne result $\gamma$ with a probability proportional to
\begin{align}
Q(\gamma){=}\exp[1{-}g^{{-}2}]\vert \gamma \vert^2.
\label{definition_Q}
\end{align}

Values of $g$ greater than 1, however, impose the use of a cut-off and a normalization term which can rapidly decrease the probability of success. This behavior is analogous to the trade-off between the cut-off in Fock space necessary to have a high fidelity and its associated low probability of success, for a physical implementation of the NLA \cite{pandey_quantum_2013}. The post-selected implementation of the NLA appears to be a very promising alternative, and has been used to perform a measurement-based entanglement-distillation \cite{chrzanowski_measurement-based_2014}. 
 
In this Letter, we use the ability to implement an NLA before a heterodyne measurement by post-selection, and present a protocol using continuous-variable quantum teleportation which can improve the properties of the channel teleported across. Under certain circumstances it can effectively transform a lossy channel to nearly an identity channel. As an application we show that the protocol can be used to distill Bell states that have been corrupted by loss. Contrary to the protocols of \cite{ralph_quantum_2011,micuda_noiseless_2012}, no physical implementation of the NLA or non-Gaussian resources are required.  

We first show that the global transformation is equivalent to an effective system composed of a noiseless amplification (or attenuation), followed by an effective Gaussian quantum channel with modified values of transmission and added noise. We then consider a regime where the teleported state effectively undergoes no loss, and an amount of thermal noise is added which can be made arbitrarily small by suitably choosing the gain of the post-selection and the EPR parameter. We obtain general results for a noisy and lossy channel, with lengthy calculations detailed in the supplementary material. 
  
In the second part, we focus on a lossy channel without thermal noise and we apply our results to a loop-hole free CHSH inequality violation and its potential application to device-independent quantum key distribution, as considered in \cite{gisin_proposal_2010}. While the CHSH inequality is not violated for the initial lossy channel, we show that teleportation with Gaussian post-selection across the same channel can lead to values of $S$ significantly larger than the classical bound of 2, with a reasonable probability of success.
  

\begin{figure}[t]
\begin{center}
\includegraphics[width= \columnwidth]{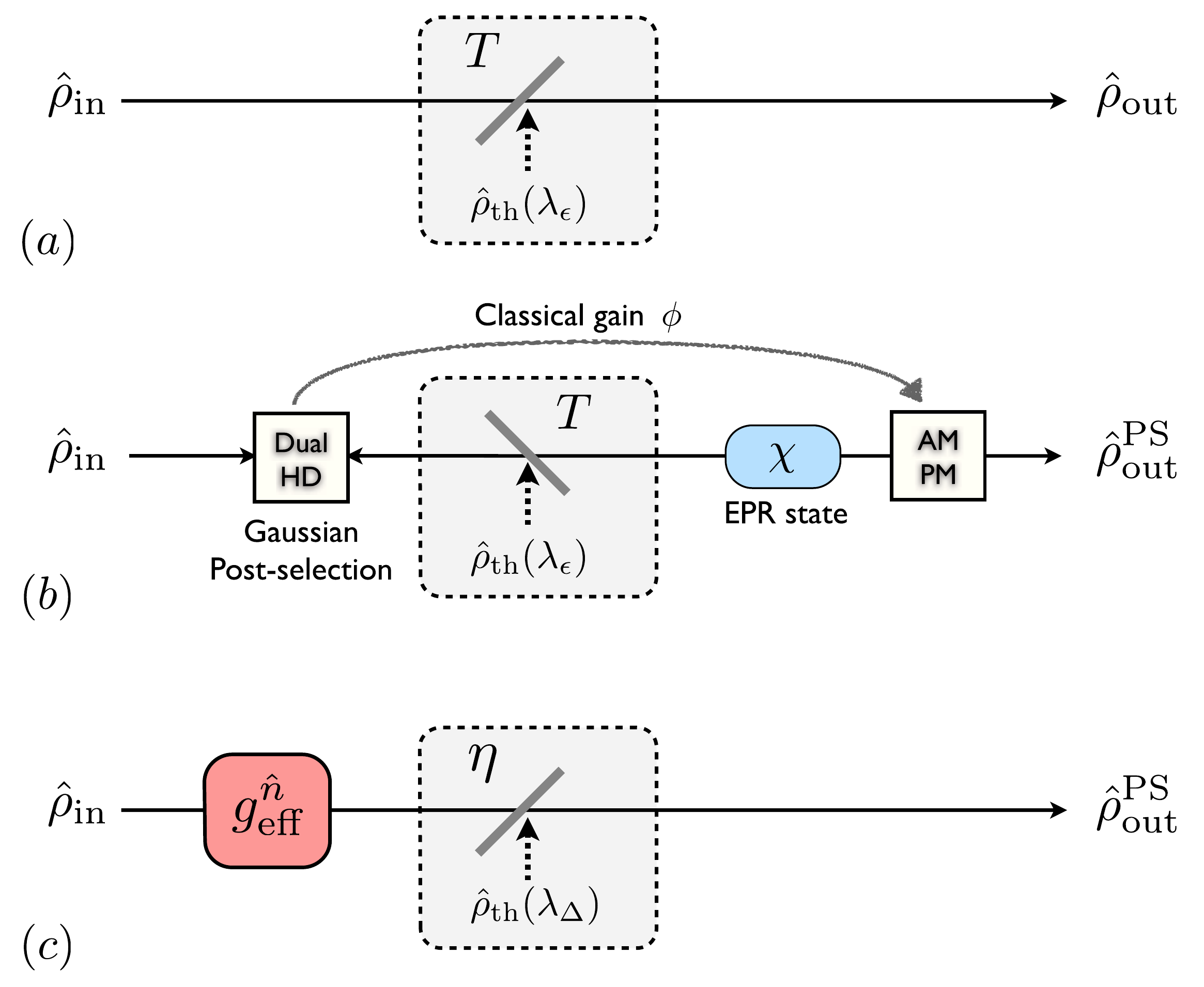} 
\caption{(a) Initial quantum channel, with a transmission $T$ and equivalent input added noise $\epsilon$, such that an input state of variance $V_{\rm in}$ is transformed to an output state of variance $V_{\rm out}{=}T(V_{\rm in}{+}\frac{1{-}T}{T}{+}\epsilon)$ \cite{Note2}. The channel can be modeled using a beam-splitter of transmission $T$ and a thermal state $\op{\rho}(\lambda_{\epsilon})$ of variance $1{+}\frac{T}{1{-}T} \epsilon$ in the second input mode. (b) Teleportation protocol with Gaussian post-selection: Bob sends one half of an EPR state through the imperfect channel and keeps the other half. Alice performs a dual-homodyne measurement (dual HD) and post-selection with the state to be transmitted, and communicates the result to Bob, who applies a corrective displacement (AM PM). (c) Equivalent effective system: the input state is noiselessly amplified (or attenuated) with a gain $g_{\rm eff}$, and sent through a quantum channel of transmission $\eta$ and added noise $\Delta$.}
\label{effective_system} 
\end{center}
\end{figure}


\paragraph{Teleportation protocol} The protocol is the following: suppose Alice prepares a state $\op{\rho}_{\rm in}$ she wants to send to Bob. The available quantum channel has a transmission $T$ and adds some thermal noise of variance $\epsilon$ referred to the input, known to Alice and Bob [Fig. \ref{effective_system} (a)]. Both agree beforehand to use the teleportation protocol and keep only the successful post-selection events. Bob prepares an EPR state $\ket{\chi}{=}\sqrt{1{-}\chi^2} \sum_{n{=}0}^\infty \chi^n \ket{n}\ket{n}$ used for the teleportation with a parameter $0{\leq}\chi{<}1$, sends one mode to Alice through the imperfect channel, and keeps the other mode. Alice performs a dual homodyne measurement with $\op{\rho}_{\rm in}$, and then performs the Gaussian post-selection with a gain $g$. Assuming that she measured the values $x$ and $p$ with the dual homodyne detection, she keeps the results with a  weighting function $Q(\frac{x{+}ip}{\sqrt{2}})$, defined by \eqref{definition_Q}.

 If the post-selection succeeds, she classically sends the results of the homodyne measurements to Bob, who applies a corrective displacement to the EPR mode he kept, with a gain $\phi$. If the post-selection fails, Alice prepares again $\op{\rho}_{\rm in}$, and the protocol is iterated until it succeeds. The protocol is depicted in Fig. \ref{effective_system} (b).

\paragraph{Effective system} We now present the main result of this Letter. As depicted in Fig. \ref{effective_system} (c), the total transformation of $\op{\rho}_{\rm in}$ can be reformulated in term of an effective system composed of an effective NLA of gain $g_{\rm eff}$, followed by a quantum channel $\mathcal{L}_{\eta,\Delta}$ of transmission $\eta$ and excess noise $\Delta$, up to a global constant factor. For the sake of simplicity, we computed the total input added noise $\chi_{\rm ch}{=}\vert(1{-}\eta)/\eta \vert{+}\Delta$, defined such that the output variance of the channel is equal to $\eta(V_{\rm in}{+}\chi_{\rm ch})$, where $V_{\rm in}$ is the variance at the input of the effective channel. 

A summary of the approach follows with the detailed calculations left to the supplementary material. Using the $P$ function it is always possible to decompose an arbitrary state into a superposition of different displacements of the input mode. In turn it is always possible to transform Fig. \ref{effective_system} (b) into an effective system for which displacements are applied instead to the EPR state, whilst the input state becomes the vacuum state. The dual homodyne detection is then equivalent to heterodyne detection of the displaced EPR state and the effect of the Gaussian post-selection is known to be equivalent to the application of an NLA before detection. This effective system containing an NLA can then be solved and is found to be equivalent to the effective channel shown in Fig. \ref{effective_system} (c). Finally numerical solutions of Fig. \ref{effective_system} (b) are calculated and found to be in good agreement with the effective channel of Fig. \ref{effective_system} (c).

Using this picture, Bob's (unnormalized) output state $\op{\rho}_{\rm out}^{\rm PS}$ reads
\begin{align}
\op{\rho}_{\rm out}^{\rm PS}= \frac{1{-}\bar{\chi}^2}{1{-}g^2 \bar{\chi}^2} g^2  \mathcal{L}_{\eta,\Delta}\left[ g_{\rm eff}^{\hat{n}} \text{ }\op{\rho}_{\rm in} \text{ } g_{\rm eff}^{\hat{n}}        \right],
\label{final_output_state_main}
\end{align}
where the effective parameters take the expressions
\begin{align}
\eta &=\frac{\big(\chi ^\star{-}g^2 \left(\phi  \left(\bar{\chi }^2{-}1\right){+}\chi ^\star\right)\big)^2}{\left(g^2 \bar{\chi
   }^2{-}1\right) \left(g^2 \left(2 \bar{\chi }^2{-}1\right){-}\bar{\chi }^2\right)}, 
   \label{expression_eta_main} \\
\chi_{\rm ch}&=\frac{1}{\eta}\frac{1{+}\lambda^2_{\rm B}{+}\lambda^2_{\rm tele}{-}3\lambda^2_{\rm B}\lambda^2_{\rm tele}}{(1{-}\lambda^2_{\rm B})(1{-}\lambda^2_{\rm tele})} {-}1,
\label{expression_chiTotCh_main} \\
g_{\rm eff}&=\sqrt{\frac{\bar{\chi}^2{-}g^2 \left(2 \bar{\chi} ^2{-}1\right)}{1{-}g^2 \bar{\chi} ^2}},
\label{definition_geff_main}
\end{align} 
with \mbox{$
\lambda^2_{\rm tele}{=} \frac{g^2 \left(\phi{-}\chi^\star\right)^2}{1{+}g^2 \left(\phi{-}\bar{\chi}{-}\chi ^\star
   \right) \left(\phi{+}\bar{\chi }{-}\chi^\star \right)}$}, \mbox{$\lambda^2_{\rm B}{=}\chi \frac{ T (\epsilon {-}2){+}2}{T \epsilon {+}2}$}, \mbox{$\chi^\star {=} \frac{2 \sqrt{T} \chi }{ 2{+}T \epsilon {-} \chi ^2 (2{+}(\epsilon{-}2)T)}$} and \mbox{$\bar{\chi}^2{=}\frac{T \left(\chi ^2 (\epsilon {-}2){-}\epsilon \right)}{\chi ^2 (T (\epsilon
   {-}2){+
}2){-}T \epsilon {-}2}$}. This results is also valid if the input state is a multimode state, with one sent through the channel.

Note that $\lambda_{B}$, $\chi^\star$ and $\bar{\chi}$ are only due to the imperfect initial channel, and do not depend on the post-selection. For a perfect channel, we have $\lambda_{B}{=}0$, and $\bar{\chi}{=}\chi^{\star}{=}\chi$. We can also obtain the effective channel for the teleportation without post-selection by taking $g{=}1$. When $g{<1}$, the probability of success of the post-selection $p_{\rm PS}$ is directly obtained by 
\begin{align}
p_{\rm PS}=\operatorname{Tr}\{\op{\rho}_{\rm out}^{\rm PS}\}. 
\label{Ppostselection}
\end{align}

\paragraph{Loss suppression} Several proposals using an NLA have been proposed to correct loss \cite{ralph_quantum_2011,micuda_noiseless_2012,blandino_noiseless_2014}, requiring a physical implementation of noiseless linear amplification. On the other hand, our loss suppression protocol requires only Gaussian post-selection. It aims to suppress the loss of the channel, while adding as little thermal noise as possible. Therefore, we consider the regime where $\chi$ and $g$ are chosen such that $\eta{=}1$, which is always possible. Using the expression \eqref{expression_eta_main} of $\eta$ with $\epsilon{=}0$ and $\phi{=}1$, this condition is satisfied for a post-selection gain $g{=}g_{\rm opt}$ given by 
 \begin{align}
g_{\rm opt}^2=\frac{(1{-}T) T \chi ^4}{    1{-}2 \sqrt{T} \chi {-}2 (1{-}T)
   \chi ^2 {+}2 \sqrt{T} \chi ^3{+} \left(1{-}T{-}T^2\right) \chi ^4}.
   \label{gOpt}
\end{align}
Using this optimal gain, the effective noise $\Delta$ takes a simple expression, given by
   \begin{align}
  \Delta=\frac{2 (1-T)\chi^2}{1-\chi^2}.
  \label{Delta_gOpt}
  \end{align}
  
As we see, the best strategy is not to use the strongest entanglement as is the case without post-selection, but rather a very weak one since $\Delta$ tends to zero when $\chi$ tends to zero. In this regime, $g_{\rm opt}$ also tends to zero. This behavior is the cornerstone of our protocol, and shows more rigorously that the effective noise can be made arbitrarily small using the post-selection. Since $\eta$ is always set to 1, the effective channel therefore tends to an identity channel. Note that we have taken a classical gain $\phi{=}1$, as this value also leads to a unit transmission regime without post-selection , where $\eta{=}\phi^2$. However, as $g$ decreases, the dependence on $\phi$ becomes less and less significant.

Without post-selection however, in the unit transmission regime the effective noise reads
\begin{align}
\Delta_{g{=}1}=\frac{2 (1{-}\sqrt{T} \chi)^2}{1{-}\chi ^2},
\end{align}
which tends to 2 units of shot noise when $\chi{\to}0$. $\Delta_{g{=}1}$ can tend to zero only if $T{=}1$ and $\chi{\to}1$, which corresponds to a perfect teleportation over a perfect channel.

Using $g_{\rm opt}$, the effective gain is given by
\begin{align}
g_{\rm eff}=\frac{\sqrt{T} \chi \big(1{-}\sqrt{T} \chi \big)}{1{-}\sqrt{T}
   \chi {+}(T{-}1) \chi ^2},
\end{align}
and also tends to zero when $\chi{\to}0$. If the input state is an eigenstate of the effective NLA, such as a Fock state, this property will only affect the probability of success of the transformation. On the other hand, if the input state is modified by the effective NLA, different strategies would need to be used to incorporate this effect.


\paragraph{ Entanglement distillation of Bell-states} We now present an application of our protocol to distill the entanglement of a maximally entangled discrete-variable state after a lossy channel. While Gaussian states cannot be distilled using only Gaussian operations \cite{eisert_distilling_2002,fiurasek_gaussian_2002,giedke_characterization_2002}, these no-go theorem do not hold for non-Gaussian states. We show that the distilled state can violate a CHSH inequality, and we characterize the entanglement using the concurrence \cite{wootters_entanglement_1998}.


 \begin{figure}[t]
\begin{center}
\includegraphics[width= \columnwidth]{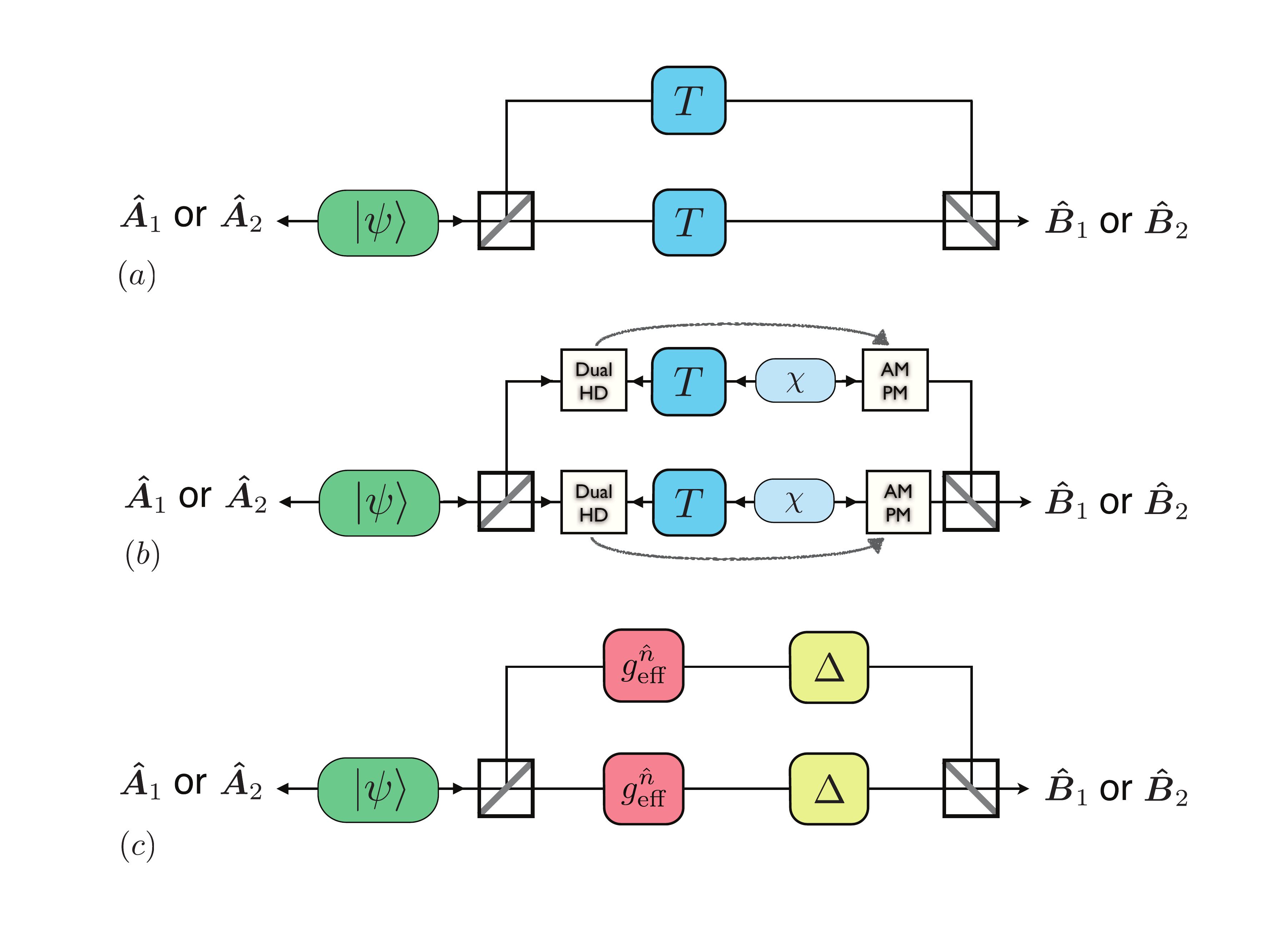}
\caption{Bell test with a maximally entangled state $\ket{\psi}$ \eqref{psi}. (a) Initial lossy channel. (b) Channel purified using teleportation and post-selection. (c) Effective picture of (b), in the unit transmission regime ($\eta{=}1$). Two channels are pictured as $\ket{\psi}$ has two independent modes $H$ and $V$. }
\label{effective_system_CHSH} 
\end{center}
\end{figure}

Let us start from a maximally polarization-entangled state
\begin{align}
\ket{\psi}=\frac{1}{\sqrt{2}}\left( \ket{1}_H\ket{1}_H+\ket{1}_V\ket{1}_V    \right),
\label{psi}
\end{align} 
following the protocol of \cite{gisin_proposal_2010}. This state could, for instance, be produced using two two-mode squeezed vacuums and a single-photon measurement. One half of $\ket{\psi}$ is sent to Alice, and the other half is sent to Bob through a lossy channel of transmission $T$. We consider two cases: either Bob directly measures the output state of the channel [Fig. \ref{effective_system_CHSH} (a)], or he uses two teleporters with post-selection to purify the modes $V$ and $H$ [Fig. \ref{effective_system_CHSH} (b)]. In both cases, Alice randomly chooses her measurement among 
\begin{align}
\op{A}_1=(\op{\sigma}_z{+}\op{\sigma}_x)/\sqrt{2}, \qquad \op{A}_2= (\op{\sigma}_z{-}\op{\sigma}_x)/\sqrt{2},
\end{align}
and Bob randomly chooses his measurement among 
\begin{align}
\op{B}_1=\op{\sigma}_z, \qquad \op{B}_2=\op{\sigma}_x,
\end{align}
as shown in Fig. \ref{effective_system_CHSH}. Each of these measurements have $\pm1$ outputs.

Alice and Bob's detectors are assumed to be unit-efficiency and photon-number-resolving. Alice will therefore always measure one photon and obtain a conclusive event. Bob, however, may obtain zero photon because of the loss, or more than one photon when using the teleportation protocol. In that case, the correlations with Alice are neglected and the measurements are discarded for simplicity, but the proportion of those inconclusive events is taken into account since we do not make the fair sampling assumption. We denote by $p_\checkmark^{\rm loss}$ (resp. $p_\checkmark^{\rm tele}$) the probability that Bob's reduced state after the initial lossy channel (resp. after the initial lossy channel corrected using the teleporter) is in the subspace $\mathcal{H}_1^B{=}\operatorname{span}\{\ket{1}_H,\ket{1}_V \}$ containing exactly one photon. 

The CHSH quantity for the conclusive events is obtained by computing the average value of the operator
\begin{align}
\op{S}&=\op{A}_1 (\op{B}_1{+}\op{B}_2)+\op{A}_2 (\op{B}_1{-}\op{B}_2)
\end{align}
when restricting Bob's state to the $\mathcal{H}_1^B$ subspace. As detailed is the supplementary material, the averaged values $S^{\rm loss}$ (resp. $S^{\rm tele}$) obtained with the lossy channel only (resp. with the teleporter) are then given by:
\begin{align}
S^{\rm loss}=p_\checkmark^{\rm loss} S_{\checkmark}^{\rm loss} =T 2\sqrt{2} \\
S^{\rm tele}=p_\checkmark^{\rm tele} S_{\checkmark}^{\rm tele}=\frac{2 \sqrt{2}}{\big(1{+}\frac{\Delta}{2}\big)^4}
\label{stele}
\end{align}
where $\Delta$ is given by \eqref{Delta_gOpt}. The concurrences $\mathcal{C}^{\rm loss}=p_\checkmark^{\rm loss} \mathcal{C}_{\checkmark}^{\rm loss}$ and $\mathcal{C}^{\rm tele}=p_\checkmark^{\rm tele} \mathcal{C}_{\checkmark}^{\rm tele}$ are obtained using the same technique.  

Note that there is no correlation between Alice and Bob when a photon is lost before Bob's detector, and therefore the correlations are not underestimated in that case. With the teleporter protocol however, higher number of photons may still have some correlations with Alice's photon, so our approach simply provides a lower bound for $S^{\rm tele}$ and $\mathcal{C}^{\rm tele}$. Note also that discarding the unsuccessful post-selections does not imply a fair sampling assumption, since Bob chooses his measurement only when the post-selection succeeds, as used in \cite{gisin_proposal_2010,brask_bell_2012}. 


 \begin{figure}[t]
\begin{center}
\subfloat[]{\includegraphics[width=0.67 \columnwidth]{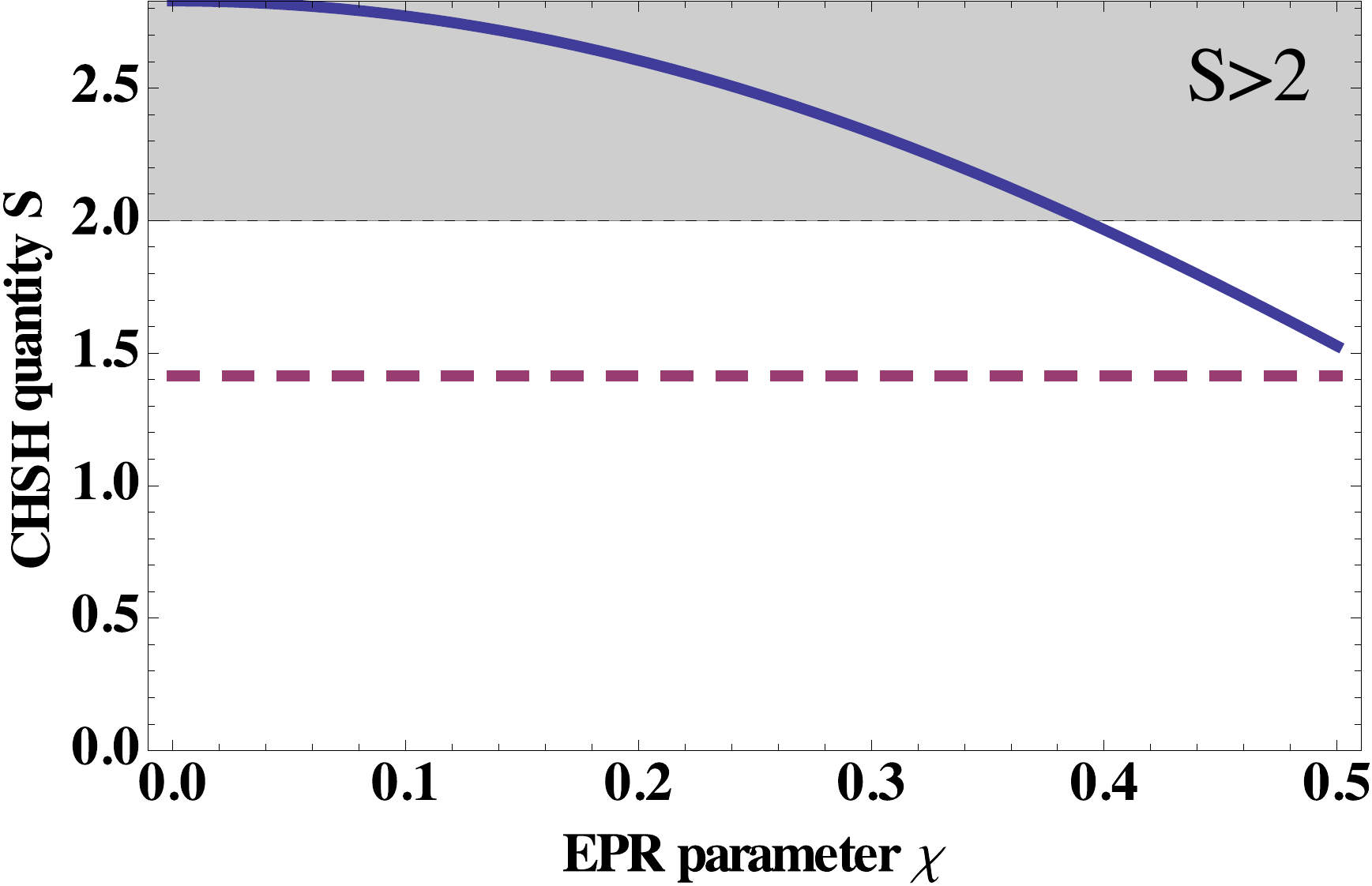}} \\
\subfloat[]{\includegraphics[width=0.67 \columnwidth]{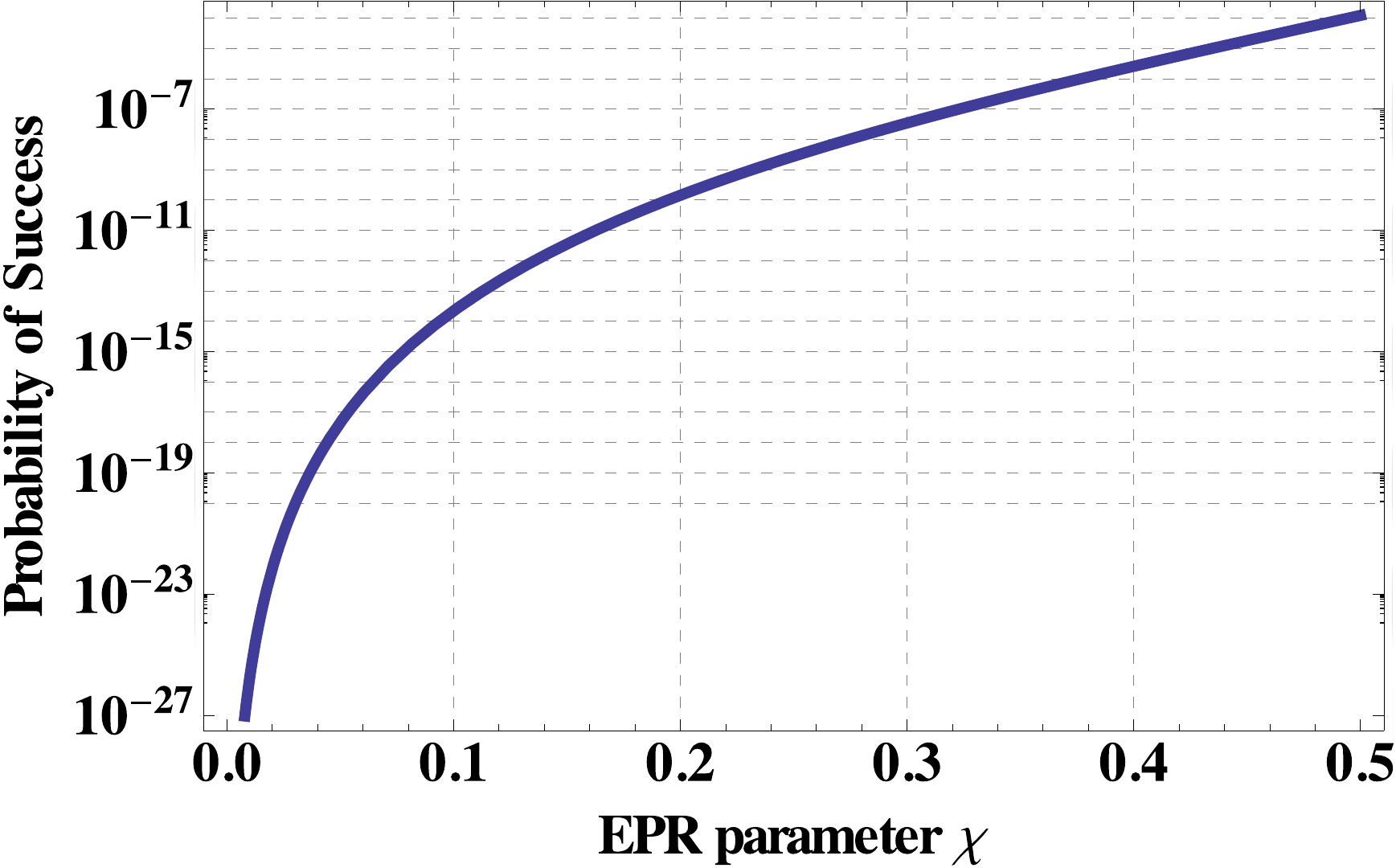}}
\caption{(a) CHSH quantities $S^{\rm tele}$ (solid line) and $S^{\rm loss}$ (dashed line). (b) Total probability of success of the two post-selection, obtained by squaring $p_{\rm PS}$ given by \eqref{Ppostselection}. For both figures, the optimal post-selection gain \eqref{gOpt} is used for each values of $\chi$, and $T{=}0.5$.}
\label{result_CHSH} 
\end{center}
\end{figure}

The CHSH quantity $S^{\rm loss}$ and $S^{\rm tele}$ are shown in Fig. \ref{result_CHSH} (a). In the limit $\chi{\to}0$, it is straightforward to see that $S^{\rm tele}{\to}2\sqrt{2}$, since the effective channel tends to the identity channel ($\Delta{\to}0$), and since the effective NLA has no effect on $\ket{\psi}$ and simply contributes to the probability of success of post-selection. Although theoretically interesting, this limit naturally corresponds to a zero probability of success. On the other hand, for non zero values of $\chi$, it is also possible to observe a violation of the CHSH inequality, with $S^{\rm tele}{>}2$, whereas $S^{\rm loss}{<}2$ as shown in Fig. \ref{result_CHSH} (a). As a concrete example, consider a transmission $T{=}0.5$, and $\chi{=}0.31$. We have $S^{\rm loss}{=}1.4$, while $S^{\rm tele}{=}2.3$. This corresponds to a probability of success of post-selection as high as $5.5{\times}10^{-8}$, as shown in Fig. \ref{result_CHSH} (b). The concurrence is also increased, from $C^{\rm loss}{=}0.5$ to $C^{\rm tele}{=}0.81$.


 \paragraph{Conclusion} Quantum teleportation with Gaussian post-selection can convert a lossy channel to an effective lossless channel with a small addition of excess thermal noise, preceded by an effective NLA with a gain smaller than 1.  This result does not depend on the input state, and is also valid for non-Gaussian entangled states.

 Certain pure states such as the Bell state considered in this Letter are not transformed by the effective NLA. However, they benefit from the effective channel improvement, which can be made arbitrarily close to an identity channel. Using this property, we have shown that our protocol can be used to distill discrete-variable entanglement and violate a CHSH inequality, which could have concrete application for device-independent quantum key distribution. 

This result is not in contradiction with the no-go theorems  \cite{niset_no-go_2009,eisert_distilling_2002,fiurasek_gaussian_2002,giedke_characterization_2002}, since we considered the purification of a non-Gaussian state. Any Gaussian state will be modified by the effective NLA, with its entanglement reduced since we considered the regime where $g{<}1$, which therefore compensates the improvement of the channel for those states. Strategies that can be used to overcome this issue will be the subject of a future publication.

\paragraph{Acknowledgment} This research was funded by the Australian Research Council Centre of Excellence for Quantum Computation and Communication Technology (Project No. CE11000102).

\newpage

\appendix

\begin{center}
\large \textbf{Supplementary Material}
\end{center}

\section{Preliminary results}
In this section, we detail the calculation of the results used in the main text. We assume that the input state to be teleported is described by a mode $\op{a}$, whereas the half of the EPR state used for the Bell measurement is described by a mode $\op{b}$. A symmetric Beam-Splitter (BS) described by an operator $\op{U}_{\rm BS}{=}\exp \left[ \theta(\op{a}^\dagger \op{b}{-}\op{a}\op{b}^\dagger)\right]$ \cite{kim_recent_2008} mixes the two modes, with $\theta{=}\pi/4$, and homodyne detections are performed on the two outputs, measuring the $\op{P}$ quadrature on the mode $\op{a}$ (noted $\op{P}_a$) and the $\op{X}$ quadrature on the mode $\op{b}$ (noted $\op{X}_b$). In the following, the subscripts $a$ and $b$ will always refer to the modes $\op{a}$ and $\op{b}$. A given outcome $(p,x)$ of the measurements is thus described by the operator
\begin{align}
\op{E}(p,x)=\bra{p}_a\bra{x}_b \op{U}_{\rm BS},
\label{operator_measurement}
\end{align} 
where $\bra{p}_a$ and $\bra{x}_b$ respectively correspond to eigenstates of the $\op{P}_a$ and $\op{X}_b$ quadratures. When the input mode $\op{a}$ is the vacuum $\ket{0}_a$, $\bra{p}_a\bra{x}_b \op{U}_{\rm BS}\ket{0}_a$ corresponds to an heterodyne measurement on the mode $\op{b}$, and it is therefore possible to emulate an NLA using post-selection \cite{fiurasek_gaussian_2012}. When the mode $\op{a}$ is not empty, however, this equivalence is not straightforward. 

\begin{figure*}[t]
\begin{center}
\includegraphics[height=11 cm]{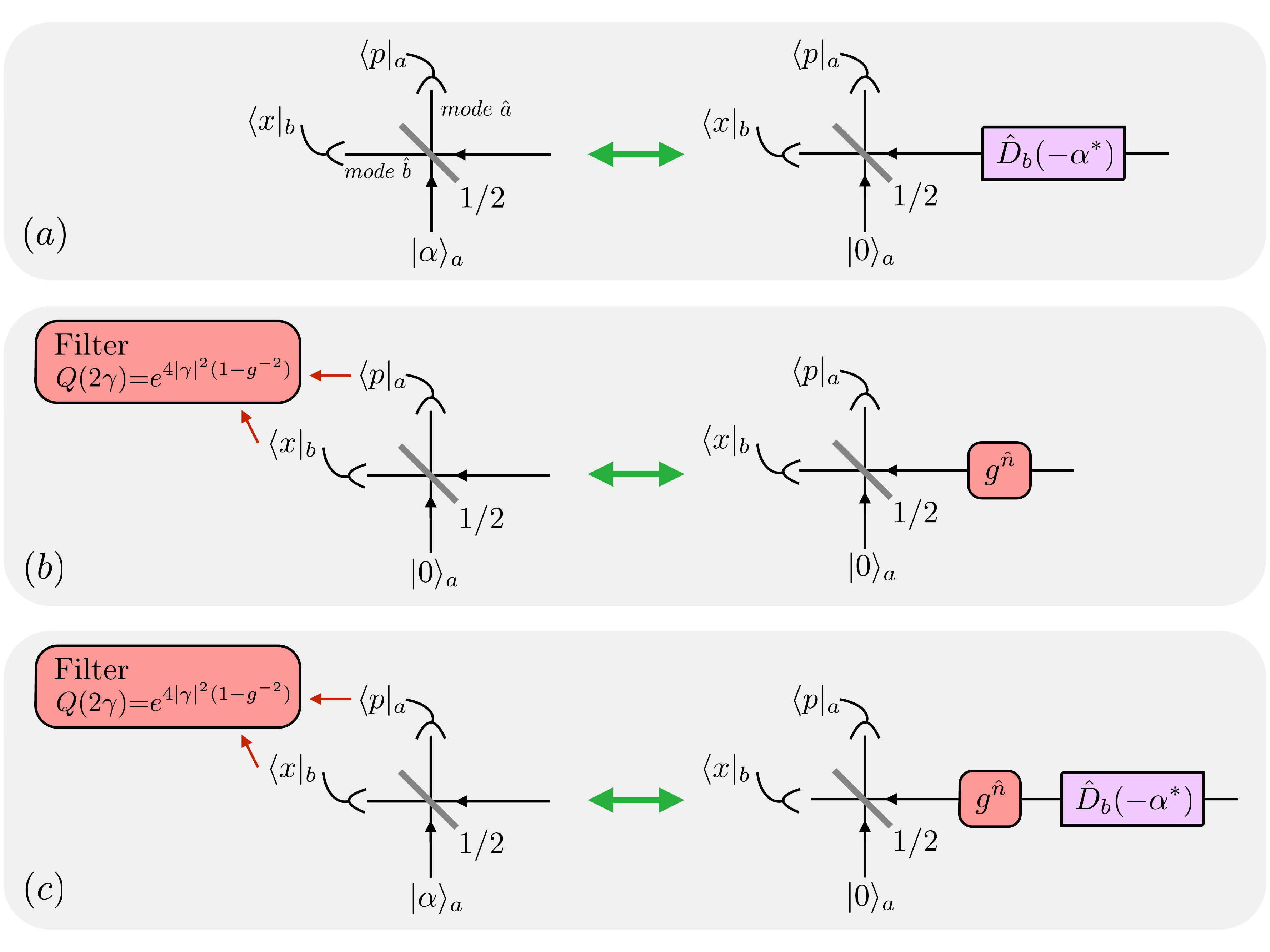} 
\caption{(a) Illustration of the relation \eqref{displaced_dualHomo}, without the phase factor. (b) Equivalence between the post-selection using the filter function $Q(2 \gamma)$ with $2 \gamma{=}\frac{x{+}ip}{\sqrt{2}}$, and an NLA. (c) Equivalences (a) and (b) used to compute the output state \eqref{rho_NLA_imperfectChannel}.}
\label{equivalences_appendix}
\end{center}
\end{figure*}

A particular case is when the mode $\op{a}$ is in a coherent state $\ket{\alpha}{=}\op{D}_a(\alpha) \ket{0}_a$, where $\op{D}_a$ is the displacement operator in the phase space. For the mode $\op{b}$, this corresponds to the operation  $\bra{p}_a\bra{x}_b \op{U}_{\rm BS}\op{D}_a(\alpha) \ket{0}_a$. As shown in the following, and illustrated in Fig. \ref{equivalences_appendix} (a), this displacement can be moved to the mode $\op{b}$,
\begin{align}
\bra{x}_b\bra{p}_a \op{U}_{\rm BS}  &\op{D}_a (\alpha) \ket{0}_a=  \nonumber \\
&\bra{x}_b\bra{p}_a \op{U}_{\rm BS} \ket{0}_a \op{D}_b ({-}\alpha^*)  e^{{-}4 i \Im(\gamma \alpha)},
\label{displaced_dualHomo}
\end{align}
 which corresponds now to an heterodyne measurement on the mode $\op{b}$ up to a phase factor,
\begin{align}
\bra{x}_b\bra{p}_a \op{U}_{\rm BS} \ket{0}_a=\frac{1}{\sqrt{2\pi}} \bra{2 \gamma},
\label{equivalence_hetero}
\end{align}
where we defined
\begin{align} 
 \gamma{=}\frac{x{+}ip}{2 \sqrt{2}}.
 \end{align}
 Using this equivalence, we can apply a Gaussian post-selection on the values $2 \gamma$ as illustrated in Fig. \ref{equivalences_appendix} (b). 

\subsubsection{Moving the displacement}

We know detail the calculation leading to \eqref{displaced_dualHomo}.
\paragraph{Simplification of the measurement operator}

 Let us first simplify the expression \eqref{operator_measurement}. According to the normalization of the vacuum noise \cite{Note2} $\op{X}_a{=}\op{a}{+}\op{a}^\dagger$, and $\op{P}_a{=}i(\op{a}^\dagger{-}\op{a})$. The infinitely squeezed states $\ket{p}_a$ and $\ket{x}_b$ can be obtained by displacing the infinitely squeezed vacuums,
\begin{subequations}
\begin{align}
\op{D}_a(ip/2) \ket{0}_{a, p}&=  \ket{p}_a,  \\
\op{D}_b(x/2) \ket{0}_{b, x}&= \ket{x}_b,
\end{align}
\end{subequations}
where $\ket{0}_{a, p}$ and $ \ket{0}_{b, x}$ are respectively the eigenstates of $\op{P}_a$ and $\op{X}_b$ with the zero value. Note that the factor 2 in the displacements comes from the relation between a coherent state and its quadratures:  $\langle\op{X} \rangle{+}i \langle\op{P} \rangle{=}2 \langle \op{a} \rangle$. The hermitian conjugate of \eqref{operator_measurement} is therefore given by
\begin{align}
\op{U}^\dagger_{\rm BS} \ket{p}_a\ket{x}_b=\op{U}^\dagger_{\rm BS} \op{D}_a(ip/2)\op{D}_b(x/2) \ket{0}_{a, p}  \ket{0}_{b, x}.
\end{align}
Inserting the identity $\op{U}_{\rm BS} \op{U}_{\rm BS}^\dagger{=}\mathbb{I}$ two times, and since $\op{U}^\dagger_{\rm BS} \op{D}_a(ip/2) \op{U}_{\rm BS} {=} \op{D}_a(ip/2\sqrt{2}) \op{D}_b(ip/2\sqrt{2})$ and $\op{U}^\dagger_{\rm BS} \op{D}_b(x/2) \op{U}_{\rm BS} {=} \op{D}_a({-}x/2\sqrt{2}) \op{D}_b(x/2\sqrt{2})$,
we finally obtain
\begin{align}
&\op{U}^\dagger_{\rm BS} \ket{p}_a\ket{x}_b= \op{D}_a(-\gamma^*)\op{D}_b(\gamma)  \op{U}_{\rm BS}^\dagger \ket{0}_{a, p} \ket{0}_{b, x}.
\end{align}

The last step is to show that $\op{U}_{\rm BS}^\dagger \ket{0}_{a, p} \ket{0}_{b, x}$ is equal to an infinitely squeezed EPR state, with the squeezed quadratures being $\op{X}_a{-}\op{X}_b$ and $\op{P}_a{+}\op{P}_b$. A (physical) single-mode squeezed vacuum along the $\op{X}$ quadrature is obtained by applying the squeezing operator $\op{S}(r){=}\exp\big[\frac{r}{2}(\op{a}^2{-}\op{a}^{\dagger 2})\big]$ on the vacuum, while a squeezed state along the $\op{P}$ quadrature is obtained using $\op{S}({-}r)$.

This leads to 
\begin{align}
\op{U}_{\rm BS}^\dagger &\op{S}_a(-r)\op{S}_b(r) \ket{0} =\exp\left[ r \op{a}^\dagger \op{b}^\dagger{-}r\op{a}\op{b} \right]\ket{0},
\label{physical_limit_homodyne_state}
\end{align}
which is a two-mode squeezed state along  $\op{X}_a{-}\op{X}_b$ and $\op{P}_a{+}\op{P}_b$. In the limit of infinite squeezing ($r{\to}\infty$), the state \eqref{physical_limit_homodyne_state} becomes a perfectly correlated EPR state 
\begin{align}
\ket{ {\rm EPR}} \propto \int \ud y \text{ } \ket{y}_a\ket{y}_b,
\end{align}
where $\ket{y}_a$ and $\ket{y}_b$ are eigenstates of $\op{X}_a$ and $\op{X}_{b}$. In conclusion, we have therefore shown that
\begin{align}
\op{U}^\dagger_{\rm BS} \ket{p}_a\ket{x_b}= \op{D}_a(-\gamma^*)\op{D}_b(\gamma) \ket{{\rm EPR}}.
\label{egalite1}
\end{align}

\paragraph{Moving the displacement}
\label{sec_preliminary_results_movDisp_2}

We now use \eqref{egalite1} to show \eqref{displaced_dualHomo}. The first step is to see the effect of an arbitrary displacement $\op{D}(\alpha)$ on an eigenstate $\ket{y}$ of $\op{X}$. Introducing $\alpha{=}u{+}iv$, since $\op{D}(\alpha){=}\op{D}(iv)\op{D}(u)\operatorname{exp}({-}i uv)$, one shows that 
\begin{align}
\op{D}(\alpha)\ket{y}=e^{iv( y{+}u)}\ket{y{+}2u},
\label{displ_sqz_state}
\end{align}
using the expression $\ket{y}{=}\frac{1}{2\sqrt{\pi}}\int \ud p \text{ } e^{-i\frac{y p}{2}}\ket{p}$ in the $\op{P}$ eigenstates basis.

Using \eqref{displ_sqz_state}, it is straightforward to show that the displacement on an infinitely squeezed EPR state reads
\begin{subequations}
\begin{align}
\op{D}_a(\alpha) \ket{{\rm EPR}}&\propto\int \ud y \op{D}_a(\alpha) \ket{y}_a \ket{y}_b \\
&=\int \ud y'  \op{D}_b({-}\alpha^*) \ket{y'}_a \ket{y'}_b 
\end{align}
\end{subequations}
where $\ket{y'}_a$ and $\ket{y'}_b$ are also eigenstates of $\op{X}_a$ and $\op{X}_b$, and therefore we obtain
\begin{align}
\op{D}_a(\alpha) \ket{{\rm EPR}}=\op{D}_b({-}\alpha^*) \ket{{\rm EPR}}.
\label{relation_EPR}
\end{align}

Consider now the state given by Eq. \eqref{egalite1}, on which we apply a displacement $\op{D}_a(\alpha)$ on the mode $\op{a}$, producing the state $\op{D}_a(\alpha)\op{D}_a(-\gamma^*)\op{D}_b(\gamma) \ket{{\rm EPR}}$. The two displacements on the mode $\op{a}$ can now be commuted using $\op{D}_a(\alpha)\op{D}_a({-}\gamma^*){=}e^{{-}2 i\Im(\alpha \gamma)}\op{D}_a({-}\gamma^*)\op{D}_a(\alpha)$.
We then use the relation \eqref{relation_EPR} to move the displacement $\op{D}_a(\alpha)$ on the mode $\op{b}$:
\begin{align}
 \op{D}_a(\alpha)&\op{D}_a(-\gamma^*)\op{D}_b(\gamma) \ket{{\rm EPR}}= \nonumber \\
& e^{-2 i\Im(\alpha \gamma)}\op{D}_a(-\gamma^*)\op{D}_b(\gamma) \op{D}_b({-}\alpha^*) \ket{{\rm EPR}}
\end{align}
Commuting  again the two displacements on the mode $\op{b}$, in order to re-obtain the expression of the heterodyne measurement, we finally obtain:
\begin{align}
\op{D}_a(\alpha)&\op{D}_a(-\gamma^*)\op{D}_b(\gamma) \ket{{\rm EPR}}= \nonumber \\
&e^{-4 i \Im(\gamma \alpha)} \op{D}_b({-}\alpha^*)\op{D}_a(-\gamma^*)\op{D}_b(\gamma) \ket{{\rm EPR}},
\end{align}
and therefore we have shown that
\begin{align}
\op{D}_a(\alpha) \op{U}^\dagger_{\rm BS} \ket{p}_a\ket{x}_b   = e^{-4 i \Im(\gamma \alpha)} \op{D}_b({-}\alpha^*)\op{U}^\dagger_{\rm BS} \ket{p}_a\ket{x}_b.
\label{moving_displacement}
\end{align}
Taking the hermitian conjugate of \eqref{moving_displacement} and changing $\alpha$ to ${-}\alpha$ leads to Eq. \eqref{displaced_dualHomo}.
 
\subsubsection{Homodyne and heterodyne measurements}
\paragraph{Equivalent heterodyne measurement} It is well known that a dual homodyne detection measuring the $\op{X}$ and $\op{P}$ quadratures corresponds to an heterodyne detection, and therefore a projection on a coherent state \cite{weedbrook_gaussian_2012}. However, one has to be careful that the different prefactors are consistent with the vacuum noise convention.

 Using the fact that $\op{U}_{\rm BS}$ can be expended as \cite{kim_recent_2008} $\op{U}_{\rm BS}{=} \exp \left[\op{a}^\dagger \op{b} \tan \theta \right]\exp\left[ {-}(\op{a}^\dagger \op{a}{-}\op{b}^\dagger \op{b}) \ln(\cos \theta) \right]{\times}$ $\exp \left[{-}\op{a}\op{b}^\dagger \tan \theta \right]$,
with $\theta{=}\pi/4$ for a symmetric BS, we first have $\op{U}_{\rm BS} \ket{0}_a=e^{\hat{a}^\dagger \hat{b}} \left(1/\sqrt{2}\right)^{\hat{n}_b}\ket{0}_a$. Using the Fock basis decomposition of $\ket{x}$, $\ket{x}{=}\sum_n \psi_n(x) \ket{n}$, where $\psi_n(x){=}\frac{e^{-x^2/4}}{(2 \pi)^{1/4} \sqrt{2^n n!}}H_n(x/\sqrt{2})$, and since the state $\ket{p}$ can be obtained with a rotation of $\pi/2$ of an eigenstate of $\op{X}$ with the same value, we have $\ket{p}=e^{i \frac{\pi}{2}\hat{n}}  \ket{x{=}p} =\sum_n \psi_n(p) i^n \ket{n}$, where $ \ket{x{=}p}$ denotes an eigenstate of $\op{X}$ with the value $p$. Using those expression to express $\bra{x}_b \bra{p}_a \op{U}_{\rm BS} \ket{0}_a$, a long but straightforward calculation leads to \eqref{equivalence_hetero}:
\begin{align}
\bra{x}_b\bra{p}_a \op{U}_{\rm BS} \ket{0}_a=\frac{1}{\sqrt{2\pi}} \bra{2 \gamma} 
\label{projection_hetero_appendix}
\end{align}
with $\gamma=\frac{x{+}ip}{2\sqrt{2}}$. 
\paragraph{Considerations on the norm} 
\label{appendix_output_norm}
When integrating over the heterodyne measurements, one has to be careful that what is measured is actually $x$ and $p$. Suppose that we have an input state $\ket{\psi}$ in mode $b$. Then, using $\int \ud x \text{ } \ket{x}\bra{x}{=}\int \ud p \text{ } \ket{p}\bra{p}{=}\mathbb{I}$, we have the following property:
 \begin{subequations}
 \begin{align}
 & \int \ud x \ud p \text{ }\bra{\psi} \bra{0}_a\op{U}_{\rm BS}^\dagger \ket{x}_b\ket{p}_a  \bra{x}_b\bra{p}_a \op{U}_{\rm BS} \ket{0}_a\ket{\psi}= \nonumber \\
  &\bra{\psi} \bra{0}_a\op{U}_{\rm BS}^\dagger \left[\int \ud x \ket{x}_b \bra{x}_b\right]\otimes\left[\int \ud p \ket{p}_a \bra{p}_a \right]\op{U}_{\rm BS} \ket{0}_a\ket{\psi} \\
  &=\braket{\psi}{\psi}\braket{0}{0}_a \\
  &=1
 \end{align}
 \end{subequations}
 This shows that, when integrating over $x$ and $p$, the norm of the state corresponds to a normalized probability density. However, by the definition of $\gamma$, we have $\ud x \ud p{=} (2 \sqrt{2})^2 \ud^2 \gamma$. Therefore, an integration over $\gamma$ must be multiplied by $ (2 \sqrt{2})^2{=}8$ to be correctly normalized.

\section{Teleportation of a coherent state}

\subsubsection{Particular homodyne outcome}
Since an arbitrary single-mode input state $\op{\rho}_{\rm in}$ to be teleported can be expressed using the $P$ function \cite{gerry_introductory_2005},
\begin{align}
\op{\rho}_{\rm in}=\int \ud^2 \alpha \text{ }P_{\rm in}(\alpha) \ket{\alpha}\bra{\alpha},
\label{rho_in_appendix}
\end{align}
and since all the transformations are linear, the knowledge of the post-selected teleportation of an arbitrary coherent state input is sufficient to obtain the post-selected teleportation of $\op{\rho}_{\rm in}$. Let us therefore consider an input state $\ket{\alpha}\bra{\alpha}$, and use Eqs. \eqref{displaced_dualHomo} and \eqref{equivalence_hetero}. We note $\op{\rho}_{\rm th}(\lambda)$ a single-mode thermal state of variance $\frac{1{+}\lambda^2}{1{-}\lambda^2}$, with a decomposition  
\begin{align}
\op{\rho}_{\rm th}(\lambda)&=\frac{1}{\pi}\frac{1{-}\lambda^2}{\lambda^2}\int \ud^2 \alpha \text{ } e^{{-}\frac{1{-}\lambda^2}{\lambda^2}\vert \alpha \vert^2}\ket{\alpha}\bra{\alpha}, \\
&=(1{-}\lambda^2)\sum_n \lambda^{2 n} \ket{n}\bra{n}.
\label{rho_th}
\end{align}

Assuming the configuration depicted in Fig. \ref{effective_system}, Bob sends on half of the EPR state $\ket{\chi}$ to Alice through the quantum channel of transmission $T$ and input excess noise $\epsilon$, and keeps the other half. The covariance matrix of this two-mode state after the channel reads \cite{weedbrook_gaussian_2012}
\begin{align}
\Gamma_{\rm AB} = \left(
\begin{array}{cc}
\Gamma_{\rm A} & C \\ 
C & \Gamma_{\rm B}
\end{array} \right),
\end{align}
where $\Gamma_{\rm A}{=}T\left(V_{\rm EPR}{+}\frac{1{-}T}{T}{+}\epsilon\right) \mathbb{I}$, $C{=}\sqrt{T(V_{\rm EPR}^2{-}1)}\mathbb{Z}$, $\Gamma_{\rm B}{=}V_{\rm EPR}\mathbb{I}$, $V_{\rm EPR}{=}\frac{1{+}\chi^2}{1{-}\chi^2}$, $\mathbb{I}{=}\operatorname{diag}(1,1)$ and $\mathbb{Z}{=}\operatorname{diag}(1,{-}1)$.
As shown above, the displacement $\op{D}(\alpha)$ creating $\ket{\alpha}\bra{\alpha}$ can be moved to Alice's EPR state mode using \eqref{displaced_dualHomo}. The Bell measurement is now equivalent to a displacement $\op{D}(-\alpha^*)$ on the EPR mode sent to Alice, followed by a projection $\bra{2 \gamma}/\sqrt{2 \pi}$, with $2 \gamma{=}(x{+}ip)/\sqrt{2}$. 

As will be explained in the following, the Gaussian post-selection is equivalent to an NLA $g^{\hat{n}}$ placed between the heterodyne detection and this displacement, up to a factor $g^2$. We can therefore use this simpler picture to compute the output state. Applying this NLA and then the displacement on $\bra{2 \gamma}$ gives 
\begin{align}
\bra{2 \gamma} g^{\hat{n}}\op{D}(-\alpha^*)=e^{\frac{1}{2}4\vert \gamma\vert ^2 (g^2{-}1)}\bra{2 g \gamma{+}\alpha^*}.
\label{action_NLA_hetero}
\end{align} 

We can therefore obtain Bob's reduced state $\op{\rho}_{\rm B}(\beta)$, for a particular heterodyne outcome, by considering the projection on a coherent state $\beta{=}\beta_x{+}i\beta_p{=}2 g \gamma{+}\alpha^*$. Its covariance matrix does not depend on $\beta$, and is given by \cite{weedbrook_gaussian_2012}
\begin{align}
\gamma_{\rm B} &=\Gamma_{\rm B}-C \left( \Gamma_{\rm A} + \mathbb{I}  \right)^{{-}1}C = V_{\rm B} \mathbb{I}
\end{align}
where 
\begin{align}
V_{\rm B} =\frac{T+V_{\rm EPR}\big(2{+}(\epsilon{-}1)T      \big)}{2+T({-}1{+}V_{\rm EPR}{+}\epsilon) }:=\frac{1{+}\lambda_{\rm B}^2}{1{-}\lambda_{\rm B}^2}.
\label{noise_vout}
\end{align}
We define 
\begin{align}
\lambda_{\rm B}=\chi \sqrt{\frac{ T (\epsilon {-}2){+}2}{T \epsilon {+}2}}
\label{lambda_B}
\end{align}
such that $V_{\rm B} {=}\frac{1{+}\lambda_{\rm B}^2}{1{-}\lambda_{\rm B}^2}$. Let us note that in the absence of loss and noise, $V_{\rm B} {=}1$ and we recover the well known fact that $\op{\rho}_{\rm B}(\beta)$ is a coherent state \cite{weedbrook_gaussian_2012}. 

Noting $\amp m{=}\sqrt{2}(\beta_x,\beta_p)^T$ \footnote{We use the convention that the variance of the vacuum quantum noise is 1.}, Bob's displacement vector $\amp{d}_{\rm out}$ reads \cite{garcia-patron_quantum_2007}
\begin{align}
\amp{d}_{\rm out}&=\sqrt{2} C \left( \Gamma_{\rm A} + \mathbb{I}  \right)^{{-}1} \amp{m} =2 \chi^\star(\beta_x,-\beta_p)^T,
\end{align}
where
\begin{align}
\chi^\star = \frac{2 \sqrt{T} \chi }{ 2{+}T \epsilon {-} \chi ^2 (2{+}(\epsilon{-}2)T)}.
\label{chi_star}
\end{align}

Since Bob's corrective displacement is unitary, the normalization of $\op{\rho}_{\rm B}(\beta)$ comes only from Alice's measurement and from the NLA, and can therefore be obtained by considering only the norm of Alice's reduced state,  which is a thermal state $\op{\rho}_{\rm th}(\bar{\chi})$ of covariance matrix $\Gamma_{\rm A}$, with 
\begin{align}
\bar{\chi}=\sqrt{\frac{T \left(\chi ^2 (\epsilon {-}2){-}\epsilon \right)}{\chi ^2 (T (\epsilon
   {-}2){+
}2){-}T \epsilon {-}2}}.
\label{chi_bar}
\end{align}
This gives a straightforward way to obtain the normalization term, using the decomposition \eqref{rho_th}:
\begin{align}
\frac{1}{2 \pi} \bra{\beta} \op{\rho}_{\rm th}(\bar{\chi}) \ket{\beta} = \frac{1}{2 \pi}(1{-}\bar{\chi}^2)e^{(\bar{\chi}^2{-}1)\vert \beta \vert ^2}
\label{normalisation_term}
\end{align}

In conclusion, after including the coefficient due to the NLA, we obtain that $\op{\rho}_{\rm B}(\beta)$ is finally given by
\begin{align}
\op{\rho}_{\rm B}(\beta)=\mathcal{N}(\alpha,\gamma)\op{D}(\chi^\star \beta^*) \op{\rho}_{\rm th}(\lambda_{\rm B})\op{D}^\dagger(\chi^\star \beta^*), 
\label{rho_beta}
\end{align}
with
\begin{align}
\mathcal{N}(\alpha,\gamma)=e^{4\vert \gamma\vert ^2 (g^2{-}1)}\frac{1}{2 \pi}(1{-}\bar{\chi}^2)e^{(\bar{\chi}^2{-}1)\vert \beta \vert ^2}.
\label{normalisation_rho_beta}
\end{align}
Note that $\chi^\star{=}\bar{\chi}{=}\chi$ for a perfect quantum channel.

\subsubsection{Average over homodyne outcome}
The next step of the protocol is that Alice communicates the value of $\gamma$ to Bob, who applies a displacement $\op{D}({-}2 g \gamma^* \phi)$ using a classical gain $\phi$. Assuming that we average over $\gamma$, the transformation of $\ket{\alpha}\bra{\alpha}$ is finally given by
\begin{align}
&\op{\sigma}_{\rm NLA}(\alpha) = \nonumber \\ 
& 8 \int \ud^2 \gamma \text{ } \mathcal{N}(\alpha,\gamma) \op{D}({-}2 g \gamma^* \phi) \op{\rho}_{\rm B}(\beta)   \op{D}({+}2 g \gamma^* \phi).
\label{sigma_NLA}
\end{align}
As explained before, a factor 8 has to be introduced since we integrate over $\gamma$ instead of on the individual homodyne measurements $x$ and $p$.

We can now see from \eqref{sigma_NLA} that the post-selection is equivalent to an NLA with an additional factor $g^2$. Following the method of \cite{fiurasek_gaussian_2012}, the post-selection is implemented by weighting each heterodyne outcome with the function $Q(2\gamma){=} e^{4\vert \gamma\vert ^2 (1{-}g^{-2})}$. With a change of variable $\gamma{=}g \bar{\gamma}$ in \eqref{sigma_NLA}, it is straightforward to show that the 
\begin{align}
 \op{\sigma}_{\rm PS}(\alpha)=g^2 \op{\sigma}_{\rm NLA}(\alpha).
   \label{total_out_state_PS_main}
\end{align}  
 
This term $g^2$ has a simple interpretation: in the post-selection case, the integration is done over the measured heterodyne outcomes, which are interpreted as being already amplified, whereas in the `physical' implementation where the NLA amplifies the measured state, the integration is done over the unamplified heterodyne outcomes. The post-selection is therefore equivalent to the action of $g^{\hat{n}{+}1}$ before the heterodyne detection.

Using the expression \eqref{rho_beta} of $\op{\rho}_{\rm B}(\beta)$, introducing $\zeta{=}2 g \gamma^*(\chi^\star{-}\phi){+}\chi^\star\alpha$ and  $\ud^2 \gamma {=}\frac{\ud^2 \zeta}{4 g^2 (\chi^\star{-}\phi)^2}$, the state $\op{\sigma}_{\rm NLA}(\alpha)$ explicitly reads
\begin{align}
&\op{\sigma}_{\rm NLA}(\alpha)= 8 \frac{1{-}\bar{\chi}^2}{2\pi} \frac{1}{4 g^2 (\chi^\star{-}\phi)^2}   \int \ud^2 \zeta \left[ \text{ }e^{\frac{4\vert\zeta{-}\chi^\star \alpha\vert^2}{4g^2(\chi^\star{-}\phi)^2}(g^2{-}1)} \right.  \nonumber \\
& \left. e^{\vert \frac{\zeta{-}\chi^\star \alpha}{\chi^\star{-}\phi}{+}\alpha \vert^2(\bar{\chi}^2{-}1) }\op{D}(\zeta)\op{\rho}_{\rm th}(\lambda_{\rm B})\op{D}({-}\zeta)  \right].
\label{rho_NLA_imperfectChannel}
\end{align}

\subsubsection{Towards the effective system}

Since the whole process including the post-selection and the teleportation is Gaussian, and acting symmetrically on both quadratures, one can expect to write $\op{\sigma}_{\rm NLA}(\alpha)$ as a displaced thermal state, with a variance that we note $V_{\rm out}$ and a displacement $G \alpha$, following the same method as \cite{blandino_noiseless_2014}. $V_{\rm out}$ will be the result of two contributions: the first one is the noise $\epsilon_{\rm B}$ due to the imperfect channel for each homodyne outcome, contained in $V_{\rm B} {:=}1{+}\epsilon_{\rm B}$ \eqref{noise_vout}, which is independent of the classical gain $\phi$. The second term, which we note $\epsilon_{\rm tele}$, results mainly from the integration over $\zeta$, and depends on $\phi$. In order to simplify the calculations, we keep separated those two contributions as 
\begin{align}
V_{\rm out}=1+\epsilon_{\rm B}+\epsilon_{\rm tele}.
\label{definition_Vout}
\end{align}
Writing $\epsilon_{\rm tele}$ as 
\begin{align}
\epsilon_{\rm tele}=\frac{1{+}\lambda^2_{\rm tele}}{1{-}\lambda^2_{\rm tele}}-1,
\end{align}
we should therefore have
\begin{align}
 \op{\sigma}_{\rm NLA}(\alpha) \propto &\frac{1}{\pi}\frac{1{-}\lambda_{\rm tele}^2}{\lambda_{\rm tele}^2}\int \ud^2 \zeta \text{ } \left[ \right.  e^{{-}\frac{1{-}\lambda_{\rm tele}^2}{\lambda_{\rm tele}^2}\vert \zeta{-}G \alpha\vert^2} \times  \nonumber \\
 &\op{D}(\zeta)\op{\rho}_{\rm th}(\lambda_{\rm B})\op{D}({-}\zeta) \left. \right].
\end{align}
This expression can be obtained if the following conditions are satisfied, respectively for the exponential coefficients of $\vert \zeta \vert^2$ and $\zeta^* \alpha$:
\begin{align}
\frac{g^2{-}1}{g^2 \left(\chi
   ^\star{-}\phi \right)^2}+\frac{\bar{\chi }^2{-}1}{\left(\chi ^\star{-}\phi \right)^2}&= - \frac{1{-}\lambda_{\rm tele}^2}{\lambda_{\rm tele}^2}  \label{condition_noisy_lambda}\\
 -\frac{\left(g^2{-}1\right) \chi ^\star}{g^2 \left(\chi
   ^\star{-}\phi \right)^2}+ \frac{\bar{\chi }^2{-}1 }{\chi ^\star{-}\phi}\Big(1{-}\frac{\chi ^\star}{\chi ^\star{-}\phi
   }\Big)&=+\frac{1{-}\lambda_{\rm tele}^2}{\lambda_{\rm tele}^2} G \label{condition_noisy_G}
\end{align}
  
Note that we do not impose the same condition for $\vert \alpha \vert^2$, since there can be an $\alpha$-dependent term coming from an effective NLA. To account for this, we introduce an additional term $\Xi$, such that
\begin{align}
{-}\frac{1{-}\lambda_{\rm tele}^2}{\lambda_{\rm tele}^2} G^2+\Xi=\left(\bar{\chi }^2{-}1\right) \Big(1{-}\frac{\chi^\star}{\chi^\star{-}\phi}\Big)^2+\frac{(g^2{-}1) \chi^{\star 2}}{g^2 \left(\chi^\star{-}\phi \right)^2} .
\label{condition_Xi}
\end{align}
Solving Eqs. \eqref{condition_noisy_lambda}   and \eqref{condition_noisy_G} gives:
\begin{align}
\lambda_{\rm tele}^2&=  \frac{g^2 \left(\phi{-}\chi^\star\right)^2}{1{+}g^2 \left(\phi{-}\bar{\chi}{-}\chi ^\star
   \right) \left(\phi{+}\bar{\chi }{-}\chi^\star \right)}  \label{definition_lambda_tele}\\
G&=\frac{\chi^\star{-}g^2 \left(\phi  \left(\bar{\chi}^2{-}1\right){+}\chi^\star\right)}{1{-}g^2
   \bar{\chi }^2}
   \label{definition_G}
\end{align}
Note that $G{=}\phi$ when $g{=}1$. Inserting $\lambda_{\rm tele}$ and $G$ in \eqref{condition_Xi} gives
\begin{align}
\Xi=\frac{(g^2{-}1)(1{-}\bar{\chi}^2)}{1{-}g^2\bar{\chi}^2}
\end{align}
Interestingly, this term does not depend on $\phi$: it is a global factor, and does not depend on $\alpha$. It can be rewritten in term of a gain $g_{\rm eff}$ such that,
\begin{align}
\Xi=g_{\rm eff}^2-1,
\end{align}
with  
\begin{align}
g_{\rm eff}=\sqrt{\frac{\bar{\chi}^2{-}g^2 \left(2 \bar{\chi} ^2{-}1\right)}{1{-}g^2 \bar{\chi} ^2}}.
\label{definition_geff}
\end{align} 
Note also that \eqref{condition_noisy_lambda} leads to 
\begin{align}
\frac{1}{g^2(\chi^\star{-}\phi)^2}=\frac{1}{1{-}g^2 \bar{\chi}^2}\frac{1{-}\lambda_{\rm tele}^2}{\lambda_{\rm tele}^2},
\end{align}
which gives the factor needed for the correct normalization of the displaced thermal state. Defining $\lambda_{\rm out}$ such that 
\begin{align}
V_{\rm out}{=}\frac{1{+}\lambda^2_{\rm out}}{1{-}\lambda^2_{\rm out}},
\label{Vout_lambdaOut}
\end{align}
one easily shows that 
\begin{align}
\lambda^2_{\rm out}=\frac{\lambda_{\rm B}^2+\lambda_{\rm tele}^2-2\lambda_{\rm B}^2\lambda_{\rm tele}^2}{1-\lambda_{\rm B}^2\lambda_{\rm tele}^2}.
\label{lambdaOut}
\end{align}

In conclusion, adding the $g^2$ term from the post-selection, we obtain the total and unnormalized transformation of a coherent state using the Gaussian post-selection:
\begin{align}
\ket{\alpha}\bra{\alpha} \to \op{\sigma}_{\rm PS}(\alpha) &=  \frac{1{-}\bar{\chi}^2}{1{-}g^2 \bar{\chi}^2} g^2  \text{ } e^{(g_{\rm eff}^2{-}1)\vert \alpha \vert^2} \times \nonumber \\
& \op{D}(G \alpha) \op{\rho}_{\rm{th}}(\lambda_{\rm out}) \op{D}({-}G \alpha)
\label{teleportation_output_final_appendix}
\end{align}

\section{Effective system}
The transformation \eqref{teleportation_output_final_appendix} can be usefully rewritten in a more general way, independent of $\alpha$, using an effective channel and an effective NLA. Let us assume that the exponential term comes from an NLA of gain $g_{\rm eff}$. This NLA would perform the transformation
\begin{align}
\ket{\alpha}\bra{\alpha} \to e^{(g_{\rm eff}^2{-}1)\vert \alpha \vert^2} \ket{g_{\rm eff} \alpha} \bra{g_{\rm eff} \alpha}.
\end{align} 
This state has still a variance equal to 1. If this state is sent through a quantum channel of transmission $\eta$ (which can be smaller than 1 for a lossy channel, or greater than 1 for a deterministic linear amplification), its mean amplitude is transformed to $\sqrt{\eta}g_{\rm eff} \alpha$. If $\eta$ is such that
\begin{align}
\sqrt{\eta}=\frac{G}{g_{\rm eff}},
\label{condition_eta}
\end{align}
the output state has the same mean amplitude as $\op{\sigma}_{\rm PS}(\alpha)$.  Defining the total equivalent input noise of the channel $\chi_{\rm ch}$, the state after the channel has the same variance as $\op{\sigma}_{\rm PS}(\alpha)$ if
\begin{align}
\eta\text{ } (1+\chi_{\rm ch})=V_{\rm out}.
\label{condition_chiCh}
\end{align} 
Note that the term 1 in the left-hand-side of \eqref{condition_chiCh} comes from the variance of the input coherent state. Using Eqs. \eqref{definition_G} and \eqref{definition_geff} in \eqref{condition_eta} leads to
\begin{align}
\eta &=\frac{\big(\chi ^\star{-}g^2 \left(\phi  \left(\bar{\chi }^2{-}1\right){+}\chi ^\star\right)\big)^2}{\left(g^2 \bar{\chi
   }^2{-}1\right) \left(g^2 \left(2 \bar{\chi }^2{-}1\right){-}\bar{\chi }^2\right)},
   \label{expression_eta}
\end{align}
 where $\chi^\star$ and $\bar{\chi}$ are respectively given by \eqref{chi_star} and  \eqref{chi_bar}. Then, using Eqs. \eqref{Vout_lambdaOut}, \eqref{lambdaOut}, \eqref{definition_lambda_tele},\eqref{lambda_B} and \eqref{expression_eta} in \eqref{condition_chiCh} leads to 
\begin{align}
\chi_{\rm ch}&=\frac{1}{\eta}\frac{1{+}\lambda^2_{\rm B}{+}\lambda^2_{\rm tele}{-}3\lambda^2_{\rm B}\lambda^2_{\rm tele}}{(1{-}\lambda^2_{\rm B})(1{-}\lambda^2_{\rm tele})} -1.
\label{expression_chiTotCh}
\end{align}

This total input noise can be interpreted as being composed of the term $\left\vert\frac{1{-}\eta}{\eta}\right \vert$ due to the loss or to the amplification, and of some excess noise $\Delta$ defined such that
\begin{align}
\chi_{\rm ch}=\left\vert\frac{1{-}\eta}{\eta}\right \vert+\Delta.
\end{align}
Note that the explicit expression of $\Delta$ is not given here due to its length.

In conclusion, we can find the parameters such that the transformation \eqref{teleportation_output_final_appendix} is given by an effective NLA of gain $g_{\rm eff}$, followed by a quantum channel of transmission $\eta$ and excess noise $\Delta$, up to a constant factor independent of the input state. Describing the quantum channel by an operator $\mathcal{L}_{\eta, \Delta}$, $\op{\sigma}_{\rm PS}(\alpha)$ reads
\begin{align}
\op{\sigma}_{\rm PS}(\alpha) =\frac{1{-}\bar{\chi}^2}{1{-}g^2 \bar{\chi}^2} g^2 \mathcal{L}_{\eta, \Delta}\left[ g_{\rm eff}^{\hat{n}} \text{ }\ket{\alpha}\bra{\alpha} \text{ } g_{\rm eff}^{\hat{n}}        \right] . 
\end{align}

The total (unnormalized) output state, obtained by post-selection and teleportation, is finally given by:
\begin{subequations}
\begin{align}
\op{\rho}_{\rm out}^{\rm PS}&= \int \ud^2 \alpha \text{ }P_{\rm in}(\alpha) \op{\sigma}_{\rm PS}(\alpha) \\
&=\frac{1{-}\bar{\chi}^2}{1{-}g^2 \bar{\chi}^2} g^2  \mathcal{L}_{\eta, \Delta}\left[ g_{\rm eff}^{\hat{n}} \text{ }\op{\rho}_{\rm in} \text{ } g_{\rm eff}^{\hat{n}}        \right]  
\label{final_output_state_appendix}
\end{align}
\end{subequations}
We stress that we kept all the normalization factors in the derivation of \eqref{final_output_state_appendix}, ant that the effective system is still valid for multimode or non-Gaussian states. Note also that Fig. \ref{effective_system} (c) assumes that $\eta{\leq}1$ for simplicity, which is the regime considered in the main text.

\section{Details of calculation of S}

The calculation of $S^{\rm loss}$ and $S^{\rm tele}$ are obtained by computing the average values of the Bell operator
\begin{align}
\op{S}&=\op{A}_1 (\op{B}_1{+}\op{B}_2)+\op{A}_2 (\op{B}_1{-}\op{B}_2) \\
	&=\frac{1}{\sqrt{2}}\left[(\op{\sigma}_z{+}\op{\sigma}_x){\otimes}(\op{\sigma}_z{+}\op{\sigma}_x){+}(\op{\sigma}_z{-}\op{\sigma}_x){\otimes}(\op{\sigma}_z{-}\op{\sigma}_x)\right]
	\label{equivalence_PS_NLA}
\end{align}
for Bob's reduced state within the subspace $\mathcal{H}_1^B{=}\operatorname{span}\{\ket{1}_H,\ket{1}_V \}$ containing exactly one photon. 

\subsubsection{Loss only}
In presence of loss only, Bob's photon can only be either transmitted with a probability $T$, either lost with a probability $1{-}T$. In the first case, Bob's reduced state is within $\mathcal{H}_1^B$, and the two-mode state is equal to $\ket{\psi}$, whereas in the second case Bob's reduced state has no component within $\mathcal{H}_1^B$. Therefore, $S_{\checkmark}^{\rm loss}{=}2 \sqrt{2}$, $p_\checkmark^{\rm loss}{=}T$, and 
\begin{align}
S^{\rm loss}=T 2\sqrt{2},
\end{align}
which is greater than 2 when $T{\geq}0.71$. Note that the loss threshold is usually found to be ${\simeq}0.83$ in the literature, but this assumes that the loss are symmetric for Alice and Bob. It also assumes that they establish a determined value for their inclusive outcomes, which leads to correlations when their measurements are both inconclusive. In our case, Alice's measurements are always conclusive, so there is no need for such a strategy. Even if Bob assigns a determined value to his inconclusive outcomes, there will not be any correlation with Alice's.

\subsubsection{Teleporter and post-selection}

We recall that the teleporter with Gaussian post-selection in the unit transmission regime is equivalent to an effective NLA of gain $g_{\rm eff}$, followed by an addition of thermal noise $\Delta$. Since the effective NLA does not modify $\ket{\psi}$, it will only affect the probability of success (Fig. \ref{effective_system}).

We write the single-photon states explicitly as two-mode states,
\begin{align}
\ket{1}_H\equiv\ket{1}_{b_1}\ket{0}_{b_2},  \qquad \ket{1}_V\equiv\ket{0}_{b_1}\ket{1}_{b_2}.
\end{align}
The modes $b_1$ and $b_2$ can be interpreted as two spatial modes, which are corrected using a teleporter with Gaussian post-selection. 

Interpreting the Gaussian noise as a random displacement \cite{ferraro_gaussian_2005}, the normalized output state of the two effective channels is given by
\begin{align}
\op{\rho}=\left(\frac{1}{\pi \Delta_{\rm ch}}\right)^2\int \ud^2 \gamma \ud^2 \beta \text{ } e^{{-}\frac{1}{\Delta_{\rm ch}}\vert\gamma\vert^2 {-}\frac{1}{\Delta_{\rm ch}}\vert\beta\vert^2} \times \nonumber \\
\op{D}_{b_1}(\gamma)\op{D}_{b_2}(\beta)  \ket{\psi}\bra{\psi}   \op{D}_{b_1}^\dagger(\gamma)\op{D}^\dagger_{b_2}(\beta), 
\end{align}
where $\Delta_{\rm ch}=\Delta/2$.

Due to the thermal noise, Bob's reduced state contains terms which do not contribute to the successful events and are not within $\mathcal{H}_1^B$. A simple way to obtain the contributing terms is to use the projector
\begin{align}
\op{\Pi}=\mathbb{I}_{\rm Alice}\otimes \Big[ \ket{1}\bra{1}_H + \ket{1}\bra{1}_V      \Big].
\end{align}
The projection of $\op{\rho}$ in $\mathcal{H}_1^B$ now writes 
\begin{align}
\op{\rho}'_\Pi=\op{\Pi} \op{\rho} \op{\Pi},
\end{align}
and can be expressed as
\begin{align}
\op{\rho}'_\Pi&=\left(\frac{1}{\pi \Delta_{\rm ch}}\right)^2\int \ud^2 \gamma \ud^2 \beta \text{ } e^{{-}\frac{1}{\Delta_{\rm ch}}\vert\gamma\vert^2 {-}\frac{1}{\Delta_{\rm ch}}\vert\beta\vert^2} \op{\rho}(\gamma,\beta),
\end{align} 
with
\begin{align}
\op{\rho}(\gamma,\beta)=\op{\Pi}\text{ }\op{D}_{b_1}(\gamma)\op{D}_{b_2}(\beta)  \ket{\psi}\bra{\psi}   \op{D}_{b_1}^\dagger(\gamma)\op{D}^\dagger_{b_2}(\beta) \text{ }\op{\Pi}.
\end{align}

Long but straightforward calculation gives
\begin{align}
p_\checkmark^{\rm tele}&=\operatorname{Tr}\{\op{\rho}'_\Pi\}, \\
&=\frac{{1+\frac{\Delta^2}{2}}}{\big(1{+}\frac{\Delta}{2}\big)^4}.
\end{align}
Normalizing $\op{\rho}'_\Pi$ and defining 
\begin{align}
\op{\rho}_\Pi=\frac{1}{p_\checkmark^{\rm tele}} \op{\rho}'_\Pi,
\end{align}
we have
\begin{align}
S_\checkmark^{\rm tele}&=\operatorname{Tr}\{\op{S}\op{\rho}_\Pi\}=\frac{2 \sqrt{2}}{1+\frac{\Delta^2}{2}}.
\end{align}

The CHSH quantity $S^{\rm tele}$ is finally given by
\begin{align}
S^{\rm tele}&=p_\checkmark^{\rm tele} S_\checkmark^{\rm tele}=\frac{2 \sqrt{2}}{\big(1{+}\frac{\Delta}{2}\big)^4},
\end{align}
which is greater than 2 when $\Delta{\leq}0.18$.

\section{Details of calculation of the concurrence}
For a two-qubit state $\op{\rho}$, let us define \cite{wootters_entanglement_1998}
\begin{align}
\op{\rho}^\star=(\op{\sigma}_y{\otimes}\op{\sigma}_y)   \op{\rho}^{T}  (\op{\sigma}_y{\otimes}\op{\sigma}_y),
\end{align}
with $\sigma_y{=}\bigl(\begin{smallmatrix}
0&{-}i\\ i&0
\end{smallmatrix} \bigr)$, and let $\{\lambda_k\}$ be the decreasing ordered eigenvalues of $\op{\rho}\op{\rho}^\star$. The concurrence $\mathcal{C}$ is given by  
\begin{align}
\mathcal{C}=\max(0,\sqrt{\lambda_1}{-}\sqrt{\lambda_2}{-}\sqrt{\lambda_3}{-}\sqrt{\lambda_4}),
\label{def_concurrence}
\end{align}
and equals 1 for a maximally entangled state and 0 for an unentangled state. We consider here a measure of entanglement linked with the CHSH inequality, that is, restricted to the successful events. Therefore, we use an `averaged' version of the concurrence $\mathcal{C}$, whose value is computed for the projection of Bob's state within $\mathcal{H}_1^B$, and weighted by the probability of success to belong to that subspace:
\begin{align}
\mathcal{C}^{\rm loss}=p_\checkmark^{\rm loss} \mathcal{C}_{\checkmark}^{\rm loss} \\
\mathcal{C}^{\rm tele}=p_\checkmark^{\rm tele} \mathcal{C}_{\checkmark}^{\rm tele}
\end{align}

\subsubsection{Loss only}

Since the concurrence of a maximally entangled state is equal to 1, we have $\mathcal{C}_{\checkmark}^{\rm loss}{=}1$, and therefore 
\begin{align}
\mathcal{C}^{\rm loss}=T.
\end{align}

\subsubsection{Teleporter and post-selection}

Following the same method as for the CHSH quantity, the concurrence $\mathcal{C}_{\checkmark}^{\rm tele}$ is obtained using the definition \eqref{def_concurrence} with the state $\op{\rho}_\Pi$. Again, long but straightforward calculation gives
\begin{align}
\mathcal{C}_{\checkmark}^{\rm tele} =  \frac{3}{\Delta ^2+2}-\frac{1}{2},
\end{align}
and therefore
\begin{align}
\mathcal{C}^{\rm tele} &=  p_\checkmark^{\rm tele}\mathcal{C}_{\checkmark}^{\rm tele} =\frac{1{-}\frac{\Delta}{2}}{(1{+}\frac{\Delta}{2})^3}.
\end{align}


\begin{thebibliography}{41}%
\makeatletter
\providecommand \@ifxundefined [1]{%
 \@ifx{#1\undefined}
}%
\providecommand \@ifnum [1]{%
 \ifnum #1\expandafter \@firstoftwo
 \else \expandafter \@secondoftwo
 \fi
}%
\providecommand \@ifx [1]{%
 \ifx #1\expandafter \@firstoftwo
 \else \expandafter \@secondoftwo
 \fi
}%
\providecommand \natexlab [1]{#1}%
\providecommand \enquote  [1]{``#1''}%
\providecommand \bibnamefont  [1]{#1}%
\providecommand \bibfnamefont [1]{#1}%
\providecommand \citenamefont [1]{#1}%
\providecommand \href@noop [0]{\@secondoftwo}%
\providecommand \href [0]{\begingroup \@sanitize@url \@href}%
\providecommand \@href[1]{\@@startlink{#1}\@@href}%
\providecommand \@@href[1]{\endgroup#1\@@endlink}%
\providecommand \@sanitize@url [0]{\catcode `\\12\catcode `\$12\catcode
  `\&12\catcode `\#12\catcode `\^12\catcode `\_12\catcode `\%12\relax}%
\providecommand \@@startlink[1]{}%
\providecommand \@@endlink[0]{}%
\providecommand \url  [0]{\begingroup\@sanitize@url \@url }%
\providecommand \@url [1]{\endgroup\@href {#1}{\urlprefix }}%
\providecommand \urlprefix  [0]{URL }%
\providecommand \Eprint [0]{\href }%
\providecommand \doibase [0]{http://dx.doi.org/}%
\providecommand \selectlanguage [0]{\@gobble}%
\providecommand \bibinfo  [0]{\@secondoftwo}%
\providecommand \bibfield  [0]{\@secondoftwo}%
\providecommand \translation [1]{[#1]}%
\providecommand \BibitemOpen [0]{}%
\providecommand \bibitemStop [0]{}%
\providecommand \bibitemNoStop [0]{.\EOS\space}%
\providecommand \EOS [0]{\spacefactor3000\relax}%
\providecommand \BibitemShut  [1]{\csname bibitem#1\endcsname}%
\let\auto@bib@innerbib\@empty
\bibitem [{\citenamefont {Bennett}\ \emph {et~al.}(1993)\citenamefont
  {Bennett}, \citenamefont {Brassard}, \citenamefont {Cr{\'e}peau},
  \citenamefont {Jozsa}, \citenamefont {Peres},\ and\ \citenamefont
  {Wootters}}]{bennett_teleporting_1993}%
  \BibitemOpen
  \bibfield  {author} {\bibinfo {author} {\bibfnamefont {C.~H.}\ \bibnamefont
  {Bennett}}, \bibinfo {author} {\bibfnamefont {G.}~\bibnamefont {Brassard}},
  \bibinfo {author} {\bibfnamefont {C.}~\bibnamefont {Cr{\'e}peau}}, \bibinfo
  {author} {\bibfnamefont {R.}~\bibnamefont {Jozsa}}, \bibinfo {author}
  {\bibfnamefont {A.}~\bibnamefont {Peres}}, \ and\ \bibinfo {author}
  {\bibfnamefont {W.~K.}\ \bibnamefont {Wootters}},\ }\href {\doibase
  10.1103/PhysRevLett.70.1895} {\bibfield  {journal} {\bibinfo  {journal}
  {Phys. Rev. Lett.}\ }\textbf {\bibinfo {volume} {70}},\ \bibinfo {pages}
  {1895} (\bibinfo {year} {1993})}\BibitemShut {NoStop}%
\bibitem [{\citenamefont {Braunstein}\ and\ \citenamefont
  {Kimble}(1998)}]{braunstein_teleportation_1998}%
  \BibitemOpen
  \bibfield  {author} {\bibinfo {author} {\bibfnamefont {S.~L.}\ \bibnamefont
  {Braunstein}}\ and\ \bibinfo {author} {\bibfnamefont {H.~J.}\ \bibnamefont
  {Kimble}},\ }\href {\doibase 10.1103/PhysRevLett.80.869} {\bibfield
  {journal} {\bibinfo  {journal} {Phys. Rev. Lett.}\ }\textbf {\bibinfo
  {volume} {80}},\ \bibinfo {pages} {869} (\bibinfo {year} {1998})}\BibitemShut
  {NoStop}%
\bibitem [{\citenamefont {Bouwmeester}\ \emph {et~al.}(1997)\citenamefont
  {Bouwmeester}, \citenamefont {Pan}, \citenamefont {Mattle}, \citenamefont
  {Eibl}, \citenamefont {Weinfurter},\ and\ \citenamefont
  {Zeilinger}}]{bouwmeester_experimental_1997}%
  \BibitemOpen
  \bibfield  {author} {\bibinfo {author} {\bibfnamefont {D.}~\bibnamefont
  {Bouwmeester}}, \bibinfo {author} {\bibfnamefont {J.-W.}\ \bibnamefont
  {Pan}}, \bibinfo {author} {\bibfnamefont {K.}~\bibnamefont {Mattle}},
  \bibinfo {author} {\bibfnamefont {M.}~\bibnamefont {Eibl}}, \bibinfo {author}
  {\bibfnamefont {H.}~\bibnamefont {Weinfurter}}, \ and\ \bibinfo {author}
  {\bibfnamefont {A.}~\bibnamefont {Zeilinger}},\ }\href {\doibase
  10.1038/37539} {\bibfield  {journal} {\bibinfo  {journal} {Nature}\ }\textbf
  {\bibinfo {volume} {390}},\ \bibinfo {pages} {575} (\bibinfo {year}
  {1997})}\BibitemShut {NoStop}%
\bibitem [{\citenamefont {Boschi}\ \emph {et~al.}(1998)\citenamefont {Boschi},
  \citenamefont {Branca}, \citenamefont {De~Martini}, \citenamefont {Hardy},\
  and\ \citenamefont {Popescu}}]{boschi_experimental_1998}%
  \BibitemOpen
  \bibfield  {author} {\bibinfo {author} {\bibfnamefont {D.}~\bibnamefont
  {Boschi}}, \bibinfo {author} {\bibfnamefont {S.}~\bibnamefont {Branca}},
  \bibinfo {author} {\bibfnamefont {F.}~\bibnamefont {De~Martini}}, \bibinfo
  {author} {\bibfnamefont {L.}~\bibnamefont {Hardy}}, \ and\ \bibinfo {author}
  {\bibfnamefont {S.}~\bibnamefont {Popescu}},\ }\href {\doibase
  10.1103/PhysRevLett.80.1121} {\bibfield  {journal} {\bibinfo  {journal}
  {Phys. Rev. Lett.}\ }\textbf {\bibinfo {volume} {80}},\ \bibinfo {pages}
  {1121} (\bibinfo {year} {1998})}\BibitemShut {NoStop}%
\bibitem [{\citenamefont {Furusawa}\ \emph {et~al.}(1998)\citenamefont
  {Furusawa}, \citenamefont {S{\o}rensen}, \citenamefont {Braunstein},
  \citenamefont {Fuchs}, \citenamefont {Kimble},\ and\ \citenamefont
  {Polzik}}]{furusawa_unconditional_1998}%
  \BibitemOpen
  \bibfield  {author} {\bibinfo {author} {\bibfnamefont {A.}~\bibnamefont
  {Furusawa}}, \bibinfo {author} {\bibfnamefont {J.~L.}\ \bibnamefont
  {S{\o}rensen}}, \bibinfo {author} {\bibfnamefont {S.~L.}\ \bibnamefont
  {Braunstein}}, \bibinfo {author} {\bibfnamefont {C.~A.}\ \bibnamefont
  {Fuchs}}, \bibinfo {author} {\bibfnamefont {H.~J.}\ \bibnamefont {Kimble}}, \
  and\ \bibinfo {author} {\bibfnamefont {E.~S.}\ \bibnamefont {Polzik}},\
  }\href {\doibase 10.1126/science.282.5389.706} {\bibfield  {journal}
  {\bibinfo  {journal} {Science}\ }\textbf {\bibinfo {volume} {282}},\ \bibinfo
  {pages} {706} (\bibinfo {year} {1998})}\BibitemShut {NoStop}%
\bibitem [{\citenamefont {Bowen}\ \emph {et~al.}(2004)\citenamefont {Bowen},
  \citenamefont {Schnabel}, \citenamefont {Lam},\ and\ \citenamefont
  {Ralph}}]{bowen_experimental_2004}%
  \BibitemOpen
  \bibfield  {author} {\bibinfo {author} {\bibfnamefont {W.~P.}\ \bibnamefont
  {Bowen}}, \bibinfo {author} {\bibfnamefont {R.}~\bibnamefont {Schnabel}},
  \bibinfo {author} {\bibfnamefont {P.~K.}\ \bibnamefont {Lam}}, \ and\
  \bibinfo {author} {\bibfnamefont {T.~C.}\ \bibnamefont {Ralph}},\ }\href
  {\doibase 10.1103/PhysRevA.69.012304} {\bibfield  {journal} {\bibinfo
  {journal} {Phys. Rev. A}\ }\textbf {\bibinfo {volume} {69}},\ \bibinfo
  {pages} {012304} (\bibinfo {year} {2004})}\BibitemShut {NoStop}%
\bibitem [{\citenamefont {Takei}\ \emph {et~al.}(2005)\citenamefont {Takei},
  \citenamefont {Aoki}, \citenamefont {Koike}, \citenamefont {Yoshino},
  \citenamefont {Wakui}, \citenamefont {Yonezawa}, \citenamefont {Hiraoka},
  \citenamefont {Mizuno}, \citenamefont {Takeoka}, \citenamefont {Ban},\ and\
  \citenamefont {Furusawa}}]{takei_experimental_2005}%
  \BibitemOpen
  \bibfield  {author} {\bibinfo {author} {\bibfnamefont {N.}~\bibnamefont
  {Takei}}, \bibinfo {author} {\bibfnamefont {T.}~\bibnamefont {Aoki}},
  \bibinfo {author} {\bibfnamefont {S.}~\bibnamefont {Koike}}, \bibinfo
  {author} {\bibfnamefont {K.-i.}\ \bibnamefont {Yoshino}}, \bibinfo {author}
  {\bibfnamefont {K.}~\bibnamefont {Wakui}}, \bibinfo {author} {\bibfnamefont
  {H.}~\bibnamefont {Yonezawa}}, \bibinfo {author} {\bibfnamefont
  {T.}~\bibnamefont {Hiraoka}}, \bibinfo {author} {\bibfnamefont
  {J.}~\bibnamefont {Mizuno}}, \bibinfo {author} {\bibfnamefont
  {M.}~\bibnamefont {Takeoka}}, \bibinfo {author} {\bibfnamefont
  {M.}~\bibnamefont {Ban}}, \ and\ \bibinfo {author} {\bibfnamefont
  {A.}~\bibnamefont {Furusawa}},\ }\href {\doibase 10.1103/PhysRevA.72.042304}
  {\bibfield  {journal} {\bibinfo  {journal} {Phys. Rev. A}\ }\textbf {\bibinfo
  {volume} {72}},\ \bibinfo {pages} {042304} (\bibinfo {year}
  {2005})}\BibitemShut {NoStop}%
\bibitem [{\citenamefont {Lee}\ \emph {et~al.}(2011)\citenamefont {Lee},
  \citenamefont {Benichi}, \citenamefont {Takeno}, \citenamefont {Takeda},
  \citenamefont {Webb}, \citenamefont {Huntington},\ and\ \citenamefont
  {Furusawa}}]{lee_teleportation_2011}%
  \BibitemOpen
  \bibfield  {author} {\bibinfo {author} {\bibfnamefont {N.}~\bibnamefont
  {Lee}}, \bibinfo {author} {\bibfnamefont {H.}~\bibnamefont {Benichi}},
  \bibinfo {author} {\bibfnamefont {Y.}~\bibnamefont {Takeno}}, \bibinfo
  {author} {\bibfnamefont {S.}~\bibnamefont {Takeda}}, \bibinfo {author}
  {\bibfnamefont {J.}~\bibnamefont {Webb}}, \bibinfo {author} {\bibfnamefont
  {E.}~\bibnamefont {Huntington}}, \ and\ \bibinfo {author} {\bibfnamefont
  {A.}~\bibnamefont {Furusawa}},\ }\href {\doibase 10.1126/science.1201034}
  {\bibfield  {journal} {\bibinfo  {journal} {Science}\ }\textbf {\bibinfo
  {volume} {332}},\ \bibinfo {pages} {330} (\bibinfo {year}
  {2011})}\BibitemShut {NoStop}%
\bibitem [{\citenamefont {Scarani}\ \emph {et~al.}(2009)\citenamefont
  {Scarani}, \citenamefont {Bechmann-Pasquinucci}, \citenamefont {Cerf},
  \citenamefont {Du{\v s}ek}, \citenamefont {L{\"u}tkenhaus},\ and\
  \citenamefont {Peev}}]{scarani_security_2009}%
  \BibitemOpen
  \bibfield  {author} {\bibinfo {author} {\bibfnamefont {V.}~\bibnamefont
  {Scarani}}, \bibinfo {author} {\bibfnamefont {H.}~\bibnamefont
  {Bechmann-Pasquinucci}}, \bibinfo {author} {\bibfnamefont {N.~J.}\
  \bibnamefont {Cerf}}, \bibinfo {author} {\bibfnamefont {M.}~\bibnamefont
  {Du{\v s}ek}}, \bibinfo {author} {\bibfnamefont {N.}~\bibnamefont
  {L{\"u}tkenhaus}}, \ and\ \bibinfo {author} {\bibfnamefont {M.}~\bibnamefont
  {Peev}},\ }\href {\doibase 10.1103/RevModPhys.81.1301} {\bibfield  {journal}
  {\bibinfo  {journal} {Rev. Mod. Phys.}\ }\textbf {\bibinfo {volume} {81}},\
  \bibinfo {pages} {1301} (\bibinfo {year} {2009})}\BibitemShut {NoStop}%
\bibitem [{\citenamefont {Caves}(1982)}]{caves_quantum_1982}%
  \BibitemOpen
  \bibfield  {author} {\bibinfo {author} {\bibfnamefont {C.~M.}\ \bibnamefont
  {Caves}},\ }\href {\doibase 10.1103/PhysRevD.26.1817} {\bibfield  {journal}
  {\bibinfo  {journal} {Phys. Rev. D}\ }\textbf {\bibinfo {volume} {26}},\
  \bibinfo {pages} {1817} (\bibinfo {year} {1982})}\BibitemShut {NoStop}%
\bibitem [{\citenamefont {Niset}\ \emph {et~al.}(2009)\citenamefont {Niset},
  \citenamefont {Fiur{\'a}{\v s}ek},\ and\ \citenamefont
  {Cerf}}]{niset_no-go_2009}%
  \BibitemOpen
  \bibfield  {author} {\bibinfo {author} {\bibfnamefont {J.}~\bibnamefont
  {Niset}}, \bibinfo {author} {\bibfnamefont {J.}~\bibnamefont {Fiur{\'a}{\v
  s}ek}}, \ and\ \bibinfo {author} {\bibfnamefont {N.~J.}\ \bibnamefont
  {Cerf}},\ }\href {\doibase 10.1103/PhysRevLett.102.120501} {\bibfield
  {journal} {\bibinfo  {journal} {Phys. Rev. Lett.}\ }\textbf {\bibinfo
  {volume} {102}},\ \bibinfo {pages} {120501} (\bibinfo {year}
  {2009})}\BibitemShut {NoStop}%
\bibitem [{\citenamefont {Eisert}\ \emph {et~al.}(2002)\citenamefont {Eisert},
  \citenamefont {Scheel},\ and\ \citenamefont
  {Plenio}}]{eisert_distilling_2002}%
  \BibitemOpen
  \bibfield  {author} {\bibinfo {author} {\bibfnamefont {J.}~\bibnamefont
  {Eisert}}, \bibinfo {author} {\bibfnamefont {S.}~\bibnamefont {Scheel}}, \
  and\ \bibinfo {author} {\bibfnamefont {M.~B.}\ \bibnamefont {Plenio}},\
  }\href {\doibase 10.1103/PhysRevLett.89.137903} {\bibfield  {journal}
  {\bibinfo  {journal} {Phys. Rev. Lett.}\ }\textbf {\bibinfo {volume} {89}},\
  \bibinfo {pages} {137903} (\bibinfo {year} {2002})}\BibitemShut {NoStop}%
\bibitem [{\citenamefont {Fiur{\'a}{\v s}ek}(2002)}]{fiurasek_gaussian_2002}%
  \BibitemOpen
  \bibfield  {author} {\bibinfo {author} {\bibfnamefont {J.}~\bibnamefont
  {Fiur{\'a}{\v s}ek}},\ }\href {\doibase 10.1103/PhysRevLett.89.137904}
  {\bibfield  {journal} {\bibinfo  {journal} {Phys. Rev. Lett.}\ }\textbf
  {\bibinfo {volume} {89}},\ \bibinfo {pages} {137904} (\bibinfo {year}
  {2002})}\BibitemShut {NoStop}%
\bibitem [{\citenamefont {Giedke}\ and\ \citenamefont
  {Ignacio~Cirac}(2002)}]{giedke_characterization_2002}%
  \BibitemOpen
  \bibfield  {author} {\bibinfo {author} {\bibfnamefont {G.}~\bibnamefont
  {Giedke}}\ and\ \bibinfo {author} {\bibfnamefont {J.~I.}~\bibnamefont
  {Cirac}},\ }\href {\doibase 10.1103/PhysRevA.66.032316} {\bibfield
  {journal} {\bibinfo  {journal} {Phys. Rev. A}\ }\textbf {\bibinfo {volume}
  {66}},\ \bibinfo {pages} {032316} (\bibinfo {year} {2002})}\BibitemShut
  {NoStop}%
\bibitem [{\citenamefont {Weedbrook}\ \emph {et~al.}(2012)\citenamefont
  {Weedbrook}, \citenamefont {Pirandola}, \citenamefont
  {Garc{\'i}a-Patr{\'o}n}, \citenamefont {Cerf}, \citenamefont {Ralph},
  \citenamefont {Shapiro},\ and\ \citenamefont
  {Lloyd}}]{weedbrook_gaussian_2012}%
  \BibitemOpen
  \bibfield  {author} {\bibinfo {author} {\bibfnamefont {C.}~\bibnamefont
  {Weedbrook}}, \bibinfo {author} {\bibfnamefont {S.}~\bibnamefont
  {Pirandola}}, \bibinfo {author} {\bibfnamefont {R.}~\bibnamefont
  {Garc{\'i}a-Patr{\'o}n}}, \bibinfo {author} {\bibfnamefont {N.~J.}\
  \bibnamefont {Cerf}}, \bibinfo {author} {\bibfnamefont {T.~C.}\ \bibnamefont
  {Ralph}}, \bibinfo {author} {\bibfnamefont {J.~H.}\ \bibnamefont {Shapiro}},
  \ and\ \bibinfo {author} {\bibfnamefont {S.}~\bibnamefont {Lloyd}},\ }\href
  {\doibase 10.1103/RevModPhys.84.621} {\bibfield  {journal} {\bibinfo
  {journal} {Rev. Mod. Phys.}\ }\textbf {\bibinfo {volume} {84}},\ \bibinfo
  {pages} {621} (\bibinfo {year} {2012})}\BibitemShut {NoStop}%
\bibitem [{\citenamefont {Ralph}(2011)}]{ralph_quantum_2011}%
  \BibitemOpen
  \bibfield  {author} {\bibinfo {author} {\bibfnamefont {T.~C.}\ \bibnamefont
  {Ralph}},\ }\href {\doibase 10.1103/PhysRevA.84.022339} {\bibfield  {journal}
  {\bibinfo  {journal} {Phys. Rev. A}\ }\textbf {\bibinfo {volume} {84}},\
  \bibinfo {pages} {022339} (\bibinfo {year} {2011})}\BibitemShut {NoStop}%
\bibitem [{\citenamefont {Mi{\v c}uda}\ \emph {et~al.}(2012)\citenamefont
  {Mi{\v c}uda}, \citenamefont {Straka}, \citenamefont {Mikova}, \citenamefont
  {Dusek}, \citenamefont {Cerf}, \citenamefont {Fiurasek},\ and\ \citenamefont
  {Jezek}}]{micuda_noiseless_2012}%
  \BibitemOpen
  \bibfield  {author} {\bibinfo {author} {\bibfnamefont {M.}~\bibnamefont
  {Mi{\v c}uda}}, \bibinfo {author} {\bibfnamefont {I.}~\bibnamefont {Straka}},
  \bibinfo {author} {\bibfnamefont {M.}~\bibnamefont {Mikova}}, \bibinfo
  {author} {\bibfnamefont {M.}~\bibnamefont {Dusek}}, \bibinfo {author}
  {\bibfnamefont {N.~J.}\ \bibnamefont {Cerf}}, \bibinfo {author}
  {\bibfnamefont {J.}~\bibnamefont {Fiurasek}}, \ and\ \bibinfo {author}
  {\bibfnamefont {M.}~\bibnamefont {Jezek}},\ }\href {\doibase
  10.1103/PhysRevLett.109.180503} {\bibfield  {journal} {\bibinfo  {journal}
  {Phys. Rev. Lett.}\ }\textbf {\bibinfo {volume} {109}},\ \bibinfo {pages}
  {180503} (\bibinfo {year} {2012})}\BibitemShut {NoStop}%
\bibitem [{Note1()}]{Note1}%
  \BibitemOpen
  \bibinfo {note} {Note that $g^{\protect \mathaccentV {hat}05E{n}}$ is
  actually unbounded only for $g{>}1$.}\BibitemShut {Stop}%
\bibitem [{\citenamefont {Ralph}\ and\ \citenamefont
  {Lund}(2008)}]{ralph_nondeterministic_2008}%
  \BibitemOpen
  \bibfield  {author} {\bibinfo {author} {\bibfnamefont {T.~C.}\ \bibnamefont
  {Ralph}}\ and\ \bibinfo {author} {\bibfnamefont {A.~P.}\ \bibnamefont
  {Lund}},\ }\href {http://arxiv.org/abs/0809.0326} {\bibfield  {journal}
  {\bibinfo  {journal} {{arXiv}:0809.0326}\ } (\bibinfo {year} {2008})},\
  \bibinfo {note} {quantum Communication Measurement and Computing Proceedings
  of 9th International Conference, Ed. A.Lvovsky, 155-160 ({AIP}, New York
  2009).}\BibitemShut {Stop}%
\bibitem [{\citenamefont {Fiur{\'a}{\v
  s}ek}(2009)}]{fiurasek_engineering_2009}%
  \BibitemOpen
  \bibfield  {author} {\bibinfo {author} {\bibfnamefont {J.}~\bibnamefont
  {Fiur{\'a}{\v s}ek}},\ }\href {\doibase 10.1103/PhysRevA.80.053822}
  {\bibfield  {journal} {\bibinfo  {journal} {Phys. Rev. A}\ }\textbf {\bibinfo
  {volume} {80}},\ \bibinfo {pages} {053822} (\bibinfo {year}
  {2009})}\BibitemShut {NoStop}%
\bibitem [{\citenamefont {Marek}\ and\ \citenamefont
  {Filip}(2010)}]{marek_coherent-state_2010}%
  \BibitemOpen
  \bibfield  {author} {\bibinfo {author} {\bibfnamefont {P.}~\bibnamefont
  {Marek}}\ and\ \bibinfo {author} {\bibfnamefont {R.}~\bibnamefont {Filip}},\
  }\href {\doibase 10.1103/PhysRevA.81.022302} {\bibfield  {journal} {\bibinfo
  {journal} {Phys. Rev. A}\ }\textbf {\bibinfo {volume} {81}},\ \bibinfo
  {pages} {022302} (\bibinfo {year} {2010})}\BibitemShut {NoStop}%
\bibitem [{\citenamefont {Walk}\ \emph
  {et~al.}(2013{\natexlab{a}})\citenamefont {Walk}, \citenamefont {Lund},\ and\
  \citenamefont {Ralph}}]{walk_nondeterministic_2013}%
  \BibitemOpen
  \bibfield  {author} {\bibinfo {author} {\bibfnamefont {N.}~\bibnamefont
  {Walk}}, \bibinfo {author} {\bibfnamefont {A.~P.}\ \bibnamefont {Lund}}, \
  and\ \bibinfo {author} {\bibfnamefont {T.~C.}\ \bibnamefont {Ralph}},\ }\href
  {\doibase 10.1088/1367-2630/15/7/073014} {\bibfield  {journal} {\bibinfo
  {journal} {New Journal of Physics}\ }\textbf {\bibinfo {volume} {15}},\
  \bibinfo {pages} {073014} (\bibinfo {year} {2013}{\natexlab{a}})}\BibitemShut
  {NoStop}%
\bibitem [{\citenamefont {Blandino}\ \emph {et~al.}(2014)\citenamefont
  {Blandino}, \citenamefont {Barbieri}, \citenamefont {Grangier},\ and\
  \citenamefont {Tualle-Brouri}}]{blandino_noiseless_2014}%
  \BibitemOpen
  \bibfield  {author} {\bibinfo {author} {\bibfnamefont {R.}~\bibnamefont
  {Blandino}}, \bibinfo {author} {\bibfnamefont {M.}~\bibnamefont {Barbieri}},
  \bibinfo {author} {\bibfnamefont {P.}~\bibnamefont {Grangier}}, \ and\
  \bibinfo {author} {\bibfnamefont {R.}~\bibnamefont {Tualle-Brouri}},\ }\href
  {http://arxiv.org/abs/1407.6798} {\bibfield  {journal} {\bibinfo  {journal}
  {{arXiv}:1407.6798 [quant-ph]}\ } (\bibinfo {year} {2014})},\ \bibinfo {note}
  {{arXiv}: 1407.6798}\BibitemShut {NoStop}%
\bibitem [{\citenamefont {Blandino}\ \emph {et~al.}(2012)\citenamefont
  {Blandino}, \citenamefont {Leverrier}, \citenamefont {Barbieri},
  \citenamefont {Etesse}, \citenamefont {Grangier},\ and\ \citenamefont
  {Tualle-Brouri}}]{blandino_improving_2012}%
  \BibitemOpen
  \bibfield  {author} {\bibinfo {author} {\bibfnamefont {R.}~\bibnamefont
  {Blandino}}, \bibinfo {author} {\bibfnamefont {A.}~\bibnamefont {Leverrier}},
  \bibinfo {author} {\bibfnamefont {M.}~\bibnamefont {Barbieri}}, \bibinfo
  {author} {\bibfnamefont {J.}~\bibnamefont {Etesse}}, \bibinfo {author}
  {\bibfnamefont {P.}~\bibnamefont {Grangier}}, \ and\ \bibinfo {author}
  {\bibfnamefont {R.}~\bibnamefont {Tualle-Brouri}},\ }\href {\doibase
  10.1103/PhysRevA.86.012327} {\bibfield  {journal} {\bibinfo  {journal} {Phys.
  Rev. A}\ }\textbf {\bibinfo {volume} {86}},\ \bibinfo {pages} {012327}
  (\bibinfo {year} {2012})}\BibitemShut {NoStop}%
\bibitem [{\citenamefont {Walk}\ \emph
  {et~al.}(2013{\natexlab{b}})\citenamefont {Walk}, \citenamefont {Ralph},
  \citenamefont {Symul},\ and\ \citenamefont {Lam}}]{walk_security_2013}%
  \BibitemOpen
  \bibfield  {author} {\bibinfo {author} {\bibfnamefont {N.}~\bibnamefont
  {Walk}}, \bibinfo {author} {\bibfnamefont {T.~C.}\ \bibnamefont {Ralph}},
  \bibinfo {author} {\bibfnamefont {T.}~\bibnamefont {Symul}}, \ and\ \bibinfo
  {author} {\bibfnamefont {P.~K.}\ \bibnamefont {Lam}},\ }\href {\doibase
  10.1103/PhysRevA.87.020303} {\bibfield  {journal} {\bibinfo  {journal} {Phys.
  Rev. A}\ }\textbf {\bibinfo {volume} {87}},\ \bibinfo {pages} {020303}
  (\bibinfo {year} {2013}{\natexlab{b}})}\BibitemShut {NoStop}%
\bibitem [{\citenamefont {Fiur{\'a}{\v s}ek}\ and\ \citenamefont
  {Cerf}(2012)}]{fiurasek_gaussian_2012}%
  \BibitemOpen
  \bibfield  {author} {\bibinfo {author} {\bibfnamefont {J.}~\bibnamefont
  {Fiur{\'a}{\v s}ek}}\ and\ \bibinfo {author} {\bibfnamefont {N.~J.}\
  \bibnamefont {Cerf}},\ }\href {\doibase 10.1103/PhysRevA.86.060302}
  {\bibfield  {journal} {\bibinfo  {journal} {Phys. Rev. A}\ }\textbf {\bibinfo
  {volume} {86}},\ \bibinfo {pages} {060302} (\bibinfo {year}
  {2012})}\BibitemShut {NoStop}%
\bibitem [{\citenamefont {Ferreyrol}\ \emph {et~al.}(2010)\citenamefont
  {Ferreyrol}, \citenamefont {Barbieri}, \citenamefont {Blandino},
  \citenamefont {Fossier}, \citenamefont {Tualle-Brouri},\ and\ \citenamefont
  {Grangier}}]{ferreyrol_implementation_2010}%
  \BibitemOpen
  \bibfield  {author} {\bibinfo {author} {\bibfnamefont {F.}~\bibnamefont
  {Ferreyrol}}, \bibinfo {author} {\bibfnamefont {M.}~\bibnamefont {Barbieri}},
  \bibinfo {author} {\bibfnamefont {R.}~\bibnamefont {Blandino}}, \bibinfo
  {author} {\bibfnamefont {S.}~\bibnamefont {Fossier}}, \bibinfo {author}
  {\bibfnamefont {R.}~\bibnamefont {Tualle-Brouri}}, \ and\ \bibinfo {author}
  {\bibfnamefont {P.}~\bibnamefont {Grangier}},\ }\href {\doibase
  10.1103/PhysRevLett.104.123603} {\bibfield  {journal} {\bibinfo  {journal}
  {Phys. Rev. Lett.}\ }\textbf {\bibinfo {volume} {104}},\ \bibinfo {pages}
  {123603} (\bibinfo {year} {2010})}\BibitemShut {NoStop}%
\bibitem [{\citenamefont {Zavatta}\ \emph {et~al.}(2011)\citenamefont
  {Zavatta}, \citenamefont {Fiur{\'a}{\v s}ek},\ and\ \citenamefont
  {Bellini}}]{zavatta_high-fidelity_2011}%
  \BibitemOpen
  \bibfield  {author} {\bibinfo {author} {\bibfnamefont {A.}~\bibnamefont
  {Zavatta}}, \bibinfo {author} {\bibfnamefont {J.}~\bibnamefont {Fiur{\'a}{\v
  s}ek}}, \ and\ \bibinfo {author} {\bibfnamefont {M.}~\bibnamefont
  {Bellini}},\ }\href {\doibase 10.1038/nphoton.2010.260} {\bibfield  {journal}
  {\bibinfo  {journal} {Nature Photon.}\ }\textbf {\bibinfo {volume} {5}},\
  \bibinfo {pages} {52} (\bibinfo {year} {2011})}\BibitemShut {NoStop}%
\bibitem [{\citenamefont {Xiang}\ \emph {et~al.}(2010)\citenamefont {Xiang},
  \citenamefont {Ralph}, \citenamefont {Lund}, \citenamefont {Walk},\ and\
  \citenamefont {Pryde}}]{xiang_heralded_2010}%
  \BibitemOpen
  \bibfield  {author} {\bibinfo {author} {\bibfnamefont {G.~Y.}\ \bibnamefont
  {Xiang}}, \bibinfo {author} {\bibfnamefont {T.~C.}\ \bibnamefont {Ralph}},
  \bibinfo {author} {\bibfnamefont {A.~P.}\ \bibnamefont {Lund}}, \bibinfo
  {author} {\bibfnamefont {N.}~\bibnamefont {Walk}}, \ and\ \bibinfo {author}
  {\bibfnamefont {G.~J.}\ \bibnamefont {Pryde}},\ }\href {\doibase
  10.1038/nphoton.2010.35} {\bibfield  {journal} {\bibinfo  {journal} {Nature
  Photon.}\ }\textbf {\bibinfo {volume} {4}},\ \bibinfo {pages} {316} (\bibinfo
  {year} {2010})}\BibitemShut {NoStop}%
\bibitem [{\citenamefont {Silberhorn}\ \emph {et~al.}(2002)\citenamefont
  {Silberhorn}, \citenamefont {Ralph}, \citenamefont {L{\"u}tkenhaus},\ and\
  \citenamefont {Leuchs}}]{silberhorn_continuous_2002}%
  \BibitemOpen
  \bibfield  {author} {\bibinfo {author} {\bibfnamefont {C.}~\bibnamefont
  {Silberhorn}}, \bibinfo {author} {\bibfnamefont {T.~C.}\ \bibnamefont
  {Ralph}}, \bibinfo {author} {\bibfnamefont {N.}~\bibnamefont
  {L{\"u}tkenhaus}}, \ and\ \bibinfo {author} {\bibfnamefont {G.}~\bibnamefont
  {Leuchs}},\ }\href {\doibase 10.1103/PhysRevLett.89.167901} {\bibfield
  {journal} {\bibinfo  {journal} {Phys. Rev. Lett.}\ }\textbf {\bibinfo
  {volume} {89}},\ \bibinfo {pages} {167901} (\bibinfo {year}
  {2002})}\BibitemShut {NoStop}%
\bibitem [{\citenamefont {Mi{\v s}ta}\ \emph {et~al.}(2010)\citenamefont {Mi{\v
  s}ta}, \citenamefont {Filip},\ and\ \citenamefont
  {Furusawa}}]{mista_continuous-variable_2010}%
  \BibitemOpen
  \bibfield  {author} {\bibinfo {author} {\bibfnamefont {L.}~\bibnamefont
  {Mi{\v s}ta}}, \bibinfo {author} {\bibfnamefont {R.}~\bibnamefont {Filip}}, \
  and\ \bibinfo {author} {\bibfnamefont {A.}~\bibnamefont {Furusawa}},\ }\href
  {\doibase 10.1103/PhysRevA.82.012322} {\bibfield  {journal} {\bibinfo
  {journal} {Phys. Rev. A}\ }\textbf {\bibinfo {volume} {82}},\ \bibinfo
  {pages} {012322} (\bibinfo {year} {2010})}\BibitemShut {NoStop}%
\bibitem [{\citenamefont {Pandey}\ \emph {et~al.}(2013)\citenamefont {Pandey},
  \citenamefont {Jiang}, \citenamefont {Combes},\ and\ \citenamefont
  {Caves}}]{pandey_quantum_2013}%
  \BibitemOpen
  \bibfield  {author} {\bibinfo {author} {\bibfnamefont {S.}~\bibnamefont
  {Pandey}}, \bibinfo {author} {\bibfnamefont {Z.}~\bibnamefont {Jiang}},
  \bibinfo {author} {\bibfnamefont {J.}~\bibnamefont {Combes}}, \ and\ \bibinfo
  {author} {\bibfnamefont {C.~M.}\ \bibnamefont {Caves}},\ }\href {\doibase
  10.1103/PhysRevA.88.033852} {\bibfield  {journal} {\bibinfo  {journal} {Phys.
  Rev. A}\ }\textbf {\bibinfo {volume} {88}},\ \bibinfo {pages} {033852}
  (\bibinfo {year} {2013})}\BibitemShut {NoStop}%
\bibitem [{\citenamefont {Chrzanowski}\ \emph {et~al.}(2014)\citenamefont
  {Chrzanowski}, \citenamefont {Walk}, \citenamefont {Assad}, \citenamefont
  {Janousek}, \citenamefont {Hosseini}, \citenamefont {Ralph}, \citenamefont
  {Symul},\ and\ \citenamefont {Lam}}]{chrzanowski_measurement-based_2014}%
  \BibitemOpen
  \bibfield  {author} {\bibinfo {author} {\bibfnamefont {H.~M.}\ \bibnamefont
  {Chrzanowski}}, \bibinfo {author} {\bibfnamefont {N.}~\bibnamefont {Walk}},
  \bibinfo {author} {\bibfnamefont {S.~M.}\ \bibnamefont {Assad}}, \bibinfo
  {author} {\bibfnamefont {J.}~\bibnamefont {Janousek}}, \bibinfo {author}
  {\bibfnamefont {S.}~\bibnamefont {Hosseini}}, \bibinfo {author}
  {\bibfnamefont {T.~C.}\ \bibnamefont {Ralph}}, \bibinfo {author}
  {\bibfnamefont {T.}~\bibnamefont {Symul}}, \ and\ \bibinfo {author}
  {\bibfnamefont {P.~K.}\ \bibnamefont {Lam}},\ }\href {\doibase
  10.1038/nphoton.2014.49} {\bibfield  {journal} {\bibinfo  {journal} {Nat
  Photon}\ }\textbf {\bibinfo {volume} {8}},\ \bibinfo {pages} {333} (\bibinfo
  {year} {2014})}\BibitemShut {NoStop}%
\bibitem [{\citenamefont {Gisin}\ \emph {et~al.}(2010)\citenamefont {Gisin},
  \citenamefont {Pironio},\ and\ \citenamefont
  {Sangouard}}]{gisin_proposal_2010}%
  \BibitemOpen
  \bibfield  {author} {\bibinfo {author} {\bibfnamefont {N.}~\bibnamefont
  {Gisin}}, \bibinfo {author} {\bibfnamefont {S.}~\bibnamefont {Pironio}}, \
  and\ \bibinfo {author} {\bibfnamefont {N.}~\bibnamefont {Sangouard}},\ }\href
  {\doibase 10.1103/PhysRevLett.105.070501} {\bibfield  {journal} {\bibinfo
  {journal} {Phys. Rev. Lett.}\ }\textbf {\bibinfo {volume} {105}},\ \bibinfo
  {pages} {070501} (\bibinfo {year} {2010})}\BibitemShut {NoStop}%
\bibitem [{Note2()}]{Note2}%
  \BibitemOpen
  \bibinfo {note} {We use the convention that the variance of the vacuum
  quantum noise is 1.}\BibitemShut {Stop}%
\bibitem [{\citenamefont {Wootters}(1998)}]{wootters_entanglement_1998}%
  \BibitemOpen
  \bibfield  {author} {\bibinfo {author} {\bibfnamefont {W.~K.}\ \bibnamefont
  {Wootters}},\ }\href {\doibase 10.1103/PhysRevLett.80.2245} {\bibfield
  {journal} {\bibinfo  {journal} {Phys. Rev. Lett.}\ }\textbf {\bibinfo
  {volume} {80}},\ \bibinfo {pages} {2245} (\bibinfo {year}
  {1998})}\BibitemShut {NoStop}%
\bibitem [{\citenamefont {Brask}\ \emph {et~al.}(2012)\citenamefont {Brask},
  \citenamefont {Brunner}, \citenamefont {Cavalcanti},\ and\ \citenamefont
  {Leverrier}}]{brask_bell_2012}%
  \BibitemOpen
  \bibfield  {author} {\bibinfo {author} {\bibfnamefont {J.~B.}\ \bibnamefont
  {Brask}}, \bibinfo {author} {\bibfnamefont {N.}~\bibnamefont {Brunner}},
  \bibinfo {author} {\bibfnamefont {D.}~\bibnamefont {Cavalcanti}}, \ and\
  \bibinfo {author} {\bibfnamefont {A.}~\bibnamefont {Leverrier}},\ }\href
  {\doibase 10.1103/PhysRevA.85.042116} {\bibfield  {journal} {\bibinfo
  {journal} {Phys. Rev. A}\ }\textbf {\bibinfo {volume} {85}},\ \bibinfo
  {pages} {042116} (\bibinfo {year} {2012})}\BibitemShut {NoStop}%
\bibitem [{\citenamefont {Kim}(2008)}]{kim_recent_2008}%
  \BibitemOpen
  \bibfield  {author} {\bibinfo {author} {\bibfnamefont {M.~S.}\ \bibnamefont
  {Kim}},\ }\href {\doibase 10.1088/0953-4075/41/13/133001} {\bibfield
  {journal} {\bibinfo  {journal} {Journal of Physics B: Atomic, Molecular and
  Optical Physics}\ }\textbf {\bibinfo {volume} {41}},\ \bibinfo {pages}
  {133001} (\bibinfo {year} {2008})}\BibitemShut {NoStop}%
\bibitem [{\citenamefont {Gerry}\ and\ \citenamefont
  {Knight}(2005)}]{gerry_introductory_2005}%
  \BibitemOpen
  \bibfield  {author} {\bibinfo {author} {\bibfnamefont {C.~C.}\ \bibnamefont
  {Gerry}}\ and\ \bibinfo {author} {\bibfnamefont {P.}~\bibnamefont {Knight}},\
  }\href@noop {} {{\emph {\bibinfo {title}
  {Introductory quantum optics}}}}\ (\bibinfo  {publisher} {Cambridge
  University Press},\ \bibinfo {address} {Cambridge, {UK}; New York},\ \bibinfo
  {year} {2005})\BibitemShut {NoStop}%
\bibitem [{\citenamefont
  {Garc{\'i}a-Patr{\'o}n}(2007)}]{garcia-patron_quantum_2007}%
  \BibitemOpen
  \bibfield  {author} {\bibinfo {author} {\bibfnamefont {R.}~\bibnamefont
  {Garc{\'i}a-Patr{\'o}n}},\ }\emph {\bibinfo {title} {Quantum Information with
  Optical Continuous Variables: from Bell Tests to Key Distribution/Information
  Quantique avec Variables Continues Optiques: des Tests de Bell {\`a} la
  Distribution de Cl{\'e}}},\ \href
  {http://theses.ulb.ac.be/ETD-db/collection/available/ULBetd-10022007-154607/}
  {Ph.D. thesis},\ \bibinfo  {school} {{ULB}} (\bibinfo {year}
  {2007})\BibitemShut {NoStop}%
\bibitem [{\citenamefont {Ferraro}\ \emph {et~al.}(2005)\citenamefont
  {Ferraro}, \citenamefont {Olivares},\ and\ \citenamefont
  {Paris}}]{ferraro_gaussian_2005}%
  \BibitemOpen
  \bibfield  {author} {\bibinfo {author} {\bibfnamefont {A.}~\bibnamefont
  {Ferraro}}, \bibinfo {author} {\bibfnamefont {S.}~\bibnamefont {Olivares}}, \
  and\ \bibinfo {author} {\bibfnamefont {M.~G.~A.}\ \bibnamefont {Paris}},\
  }\href {http://arxiv.org/abs/quant-ph/0503237} {\bibfield  {journal}
  {\bibinfo  {journal} {{arXiv}:quant-ph/0503237}\ } (\bibinfo {year}
  {2005})},\ \bibinfo {note} {(Bibliopolis, Napoli, 2005) {ISBN}
  88-7088-483-X}\BibitemShut {NoStop}%
\end{thebibliography}
\end{document}